\documentclass[twocolumn]{autart}

\usepackage{amssymb}
\usepackage{mathrsfs}
\usepackage{amsmath}
\usepackage{mathtools}
\usepackage{enumerate}
\usepackage[dvipsnames]{xcolor}
\usepackage[noadjust]{cite}
\usepackage{indentfirst}
\usepackage{graphicx} 
\graphicspath{{./fig/}}
\usepackage[font=footnotesize]{caption}
\usepackage{subcaption}
\usepackage[numbers]{natbib}

\usepackage{tikz}
\usepackage{pgfplots}
\usetikzlibrary{arrows,arrows.meta}
\pgfplotsset{compat=1.9}
\usepackage[breaklinks=true, colorlinks, bookmarks=true]{hyperref}

\newtheorem{theorem}{Theorem}[section]
\newtheorem{lemma}[theorem]{Lemma}
\newtheorem{definition}[theorem]{Definition}
\newtheorem{corollary}[theorem]{Corollary}

\newtheorem{problem}{Problem}
\newtheorem{remark}[theorem]{Remark}
\newtheorem{assumption}{Assumption}
\newtheorem{example}[theorem]{Example}

\newtheorem{proof}[theorem]{Proof}

\newcommand{\setdef}[2]{\{#1 \,| \, #2\}}

\newcommand{\nom}{\operatorname{nom}}
\newcommand{\oprocendsymbol}{\hbox{$\bullet$}}
\newcommand{\oprocend}{\relax\ifmmode\else\unskip\hfill\fi\oprocendsymbol}

\newcommand{\longthmtitle}[1]{\mbox{}\textup{\textbf{(#1).}}}

\newcommand{\stkout}[1]{\ifmmode\text{\sout{\ensuremath{#1}}}\else\sout{#1}\fi}

\definecolor{paleGreen}{rgb}{.3, .7, .3}

\newcommand{\co}[1]{\operatorname{co}(#1)}

\newcommand{\until}[1]{\{1,\dots,#1\}}

\DeclareMathOperator*{\argmin}{arg\,min}

\newcommand{\interior}[1]{\operatorname{int}(#1)}
\newcommand{\closure}[1]{\operatorname{cl}\! \left ( #1 \right )}

\newcommand{\real}{\mathbb{R}}

\newcommand{\Dc}{\mathcal{D}}
\newcommand{\Ac}{\mathcal{A}}

\newcommand{\Cc}{\mathcal{C}}
\newcommand{\Xc}{\mathcal{X}}
\newcommand{\Yc}{\mathcal{Y}}
\newcommand{\Nc}{\mathcal{N}}
\newcommand{\Mc}{\mathcal{M}}
\newcommand{\Sc}{\mathcal{S}}
\newcommand{\Uc}{\mathcal{U}}
\newcommand{\Jc}{\mathcal{J}}
\newcommand{\Ic}{\mathcal{I}}

\newcommand{\Lc}{\mathcal{L}}
\newcommand{\dt}{\frac{d}{dt}}
\renewcommand{\theenumi}{(\roman{enumi})}

\newcommand{\uact}{u_{\operatorname{act}}}
\newcommand{\uall}{u_{\operatorname{all}}}
\newcommand{\uadp}{u_{\operatorname{adp}}}
\newcommand{\alphaadp}{\alpha_{\operatorname{adp}}}
\newcommand{\Madp}{M_{\operatorname{adp}}}

\allowdisplaybreaks
\begin{document}
\begin{frontmatter}
  % \runtitle{Insert a suggested running title}  % Running title for regular 
  
  \title{Safety-Critical Control of Discontinuous Systems with
    Nonsmooth Safe
    Sets\thanksref{footnoteinfo}} % Title, preferably not more
  
  \thanks[footnoteinfo]{This paper was not presented at any IFAC
    meeting. Corresponding author Mohammed Alyaseen. All authors are
    with the Contextual Robotics Institute, University of California
    San Diego. M. Alyaseen is also affiliated with Kuwait University
    as a holder of a scholarship. This work was partially supported by
    AFOSR Award FA9550-23-1-0740.}

  \author{Mohammed Alyaseen}\ead{malyasee@ucsd.edu} \quad % Add the
  \author{Nikolay Atanasov}\ead{natanasov@ucsd.edu}
  \quad % e-mail address
  \author{Jorge Cortes}\ead{cortes@ucsd.edu} % (ead) as shown
  
  \address{University of California San Diego, 9500 Gilman Dr, La
    Jolla, CA 92093, USA} % Please supply
  
  
  % keyword list or with the help of the Automatica keyword wizard
  
  \begin{abstract}
    This paper studies the design of controllers for discontinuous
    dynamics that ensure the safety of non-smooth sets. The safe set
    is represented by arbitrarily nested unions and intersections of
    0-superlevel sets of differentiable functions.  We show that any
    optimization-based controller that satisfies only the point-wise
    active safety constraints is generally un-safe, ruling out the
    standard techniques developed for safety of continuous dynamics.
    This motivates the introduction of the notion of transition
    functions, which allow us to incorporate even the inactive safety constraints without falling into unnecessary conservatism. 
    % \NA{I find this sentence mysterious. 
    % 1. Perhaps, there is a more clear way of introducing the concept. 
    % 2. Also, is the name transition function the best choice given that it is already in use for several things in control theory.
    % For example, ``We introduce the notion of slackness functions, which relax a subset of the safety constraints, allowing system trajectories to transition among components of the nonsmooth safe set without endangering safety.'' I see that later you have good justification for using "transition function" so feel free to keep the name but do re-write this sentence.
    % 3. In the previous, sentence we mentioned that satisfying active constraints is insufficient. But here we are talking about relaxing constraints. It would be good to make the connection better, i.e., explain that instead of enforcing active constraints, we allow the non-active constraints to have an influence on the closed-loop system dynamics through the slackness function. We have a nice explanation in the contributions statement. We need some of it here.}.
    These functions allow system trajectories to
    leave a component of the nonsmooth safe set to transition to a
    different one.  The resulting controller is then defined as the
    solution to a convex optimization problem, which we show is
    feasible and continuous wherever the system dynamics is
    continuous. We illustrate the effectiveness of the proposed design
    approach in a multi-agent reconfiguration control problem.
  \end{abstract}
\end{frontmatter}

\section{Introduction}

Safety-critical control is a fundamental problem in numerous domains
including autonomous driving, power and transportation systems,
robotics, and even mitigation of epidemic spreading.  This paper
tackles the synthesis of easy-to-compute controllers for the objective of safety. Specifically, we address the problem of safety critical control to render a desired nonsmooth set safe under the trajectories of discontinuous
dynamics.
% \NA{I suggest splitting this sentence in two, i.e., first explain that we tackle synthesis of safe controllers and then explain that unlike the majority of existing work we consider nonsmooth safe sets and systems with discontinuous dynamics.} 
Discontinuous dynamical systems are convenient for modeling
a wide range of behaviors, including motion subject to Coulomb
friction, systems with abrupt changes in forces, and dynamics of
multi-agent systems under connectivity maintenance or collision
avoidance constraints. Likewise, the flexibility enabled by nonsmooth
safe sets allows us to encode safety requirements that cannot be
expressed merely by a differentiable function, such as keeping the
system evolution in or out of a set with nonsmooth boundary, or in
some nonsmooth set and out of another simultaneously. Allowing
nonsmoothness is also motivated by the fact that dealing with multiple
constraints, some of which are active in some regions while different
ones are active in others, can be represented as an overall nonsmooth
safety requirement.
%
% \marginJC{More examples here. Could we also say ``dealing with
% multiple constraints at once''? (where some are active in some
% regions, and different ones are active in other regions, resulting
% in an overall not-differentiable behavior)?}
%
The unique challenge tackled here is constructing an explicit,
provably feasible, sufficiently regular, optimization-based safe
controller for discontinuous systems with general non-smooth sets
defined by nested unions and intersections of smooth component sets.

\emph{Literature review:} Safety-critical control is often formulated
mathematically as a control design problem to achieve forward
invariance of a desired set of the state space. The notion of control
barrier function (CBF) has gained popularity as a tool for invariance
control due to its
versatility~\cite{ADA-SC-ME-GN-KS-PT:19,ADA-XX-JWG-PT:17,WX-CGC-CB:23}. Broadly
speaking, a CBF is a function for which every system state in its
0-superlevel set admits an input value that does not allow the
dynamics to leave the superlevel set instantaneously. For convenience,
we refer to this point-wise condition as the CBF condition. Achieving
control invariance via CBFs is a twofold undertaking.  First, one must
find a function for which the CBF condition holds.  Second, one must
design a controller with adequate regularity properties that keeps the
the superlevel set of the function invariant. These two problems are
in general distinct. Even if a differentiable CBF is available for
continuous dynamics, there might not exist a continuous controller
rendering its superlevel set invariant, cf. \cite{MA-NA-JC:23-tac}.
% In this work we deal with the second problem for the case of
% discontinuous dynamics and non-smooth CBFs.  Namely, we construct an
% online-calculable controller that renders a non-smooth set
% controlled-invariant under discontinuous dynamics.

In the classical setting, a CBF is a differentiable function whose
0-superlevel set is made control-invariant by employing a Lipschitz
controller that satisfies the CBF
condition \cite{ADA-SC-ME-GN-KS-PT:19}. A widely used technique for
constructing such a controller is quadratic programming (QP),
constrained by the CBF condition, which is linear in the input for
control-affine dynamics \cite{ADA-XX-JWG-PT:17}. The differentiability
requirement of the CBF stems from the use of its gradient in the CBF
condition, while the Lipschitz requirement on the dynamics and
controller ensures existence and uniqueness of solutions for the
closed-loop system \cite{HK:02}. This excludes discontinuous dynamics
and non-smooth safe sets, which can be limiting in many applications. 
However, these two structural assumptions are not
indispensable, and indeed many works have explored the extension of
CBF techniques beyond them.  In one line of work, safety conditions
are developed for autonomous hybrid systems and non-smooth barrier
functions \cite{MM-RGS:21}. These conditions extend to systems modeled
by differential inclusions with control inputs as
in \cite{MG-AI-RGS-WED:22f,AI-MG-RGS-WED:24}. The
work \cite{MG-AI-RGS-WED:22f} provides general conditions for the
existence of a continuous safe controller but does not provide an
explicit one. Instead, \cite{AI-MG-RGS-WED:24} gives an explicit
optimization-based continuous safe controller for non-smooth safe sets
defined as an intersection of superlevel sets of differentiable
functions. % This work further provides a sum-of-squares method to
% generate the safe sets for a class of systems.
The work \cite{PG-JC-ME:17-csl} gives sufficient conditions for safety
of differential inclusions and general non-smooth barrier functions,
and designs an optimization-based controller for sets defined by the
intersection of multiple superlevel sets. This controller is safe
assuming the optimization is feasible but no conditions are provided
for the latter.  These methods have been successfully applied for
multi-agent network connectivity \cite{PO-BC-LS-JC:23-auto}. Other
works \cite{JU-KG-DP:20} are not limited to continuous controllers and
instead develop Lebesgue-measurable controllers for non-smooth CBFs
whose superlevel sets are given as an intersection of multiple
differentiable functions. The works mentioned hitherto give explicit
safe controllers only for non-smooth sets defined as intersections,
that is, conjunctive safety constraints. When the safe set is given as
a union of component sets, the safety constraint on the controller
changes depending on the component set in which the current state
lies, leading to an optimization with disjunctive constraints. Such
problems are equivalent to mixed-integer optimization
programs \cite{IEG:02} and thus are more challenging to deal with.
% not as straightforward to approach
% as those with conjunctive constraints. Hence, such problems are not
% natural extensions of those with conjunctive constraints.
Disjunctive safety constraints appear in another line of work, that of
enforcing signal temporal logic (STL) specifications as barrier
functions. The work \cite{LL-DVD:20} models conjunctive signal
temporal tasks for Lipschitz dynamics as non-smooth barrier functions
and provides a generally discontinuous optimization-based controller
to enforce such constraints. The works \cite{AW-DVD:22,AZ-HGT:22} do
the same for logical specifications including disjunction. Although
the STL-induced CBF is more general than mere time-independent CBFs,
the controller developed in those works is generally discontinuous,
even with the assumed Lipschitz dynamics, and is provably un-safe for
discontinuous systems, as we establish later in the paper.

% ancla
\emph{Contributions:} We consider the problem of constructing 
an explicit, provably feasible, optimization-based safe controller
for discontinuous systems with non-smooth sets defined by nested
unions and intersections of smooth sets.  We start by considering the
\emph{active-component} QP controller, an optimization-based
controller satisfying only the point-wise active safety constraints.
We show that this controller is generally un-safe for discontinuous
systems and nonsmooth safe sets, thereby excluding the possibility of
using the techniques developed for safe control of continuous dynamics
in our setting. We leverage our analysis of the reasons behind this
failure to construct a new QP controller that overcomes them.  We do
this by considering all safety constraints, even those that are
inactive at the point at which the control is computed, to introduce
the \emph{all-components} QP controller. The
%\NA{I suggest rewriting this sentence to explain that we introduce the idea of slackness functions to allow all constraints to influence the control design, even before they are active.}
inclusion of all safety
constraints allows the inactive constraints to influence the control without not limiting for a wide range safe sets because of the
novel concept of \emph{transition function}, which modify the standard
safety conditions to facilitate their satisfaction for the inactive
constraints, thereby avoiding conservatism.  The inclusion of all
safety constraints ensures safety while enabling the continuity of the
optimization-based controller.
%\NA{The explanation here is good. We need some of this in the abstract.} 
Specifically, we show that our proposed controller is continuous wherever the dynamics are continuous
even at points of nonsmoothness of the safe set.  We prove the
feasibility of the all-components QP controller for any safe set given
by nested intersections and unions of smooth sets.  Notice that since
we allow representing our safe sets as unions, the safety conditions
include disjunctive constraints. However, instead of solving the
problem using disjunctive optimization, we develop the all-components
QP controller which we show is feasible whenever the disjunctive
program is. This QP representation of our controller is what deems it
continuous even at the points where the disjunctive safety constraints
discontinuously change. Our last contribution shows how the design
parameters of our all-components QP controller can be chosen
adaptively dependent on the state. This results in the introduction of
the \emph{all-components adaptive} QP controller. Finally, we
demonstrate the efficacy of our proposed controller by applying it to
control the motion of multi-agents with mixed safety specifications.
%
% \marginJC{The names we use later: ``all-components controller'',
% etc., should appear also in the statement of contributions, no?
% Also, why no mention to design parameters, determining them
% adaptively in ``all components adaptive controller''?}
%

\emph{Notation:} For $x \in \real^n$, we define
$B_\epsilon(x) \coloneqq \{y \in \real^n \;|\; \|x -
y\|<\epsilon\}$. For a set $\Sc \subseteq \real^n$,
$B_\epsilon(\Sc) \coloneqq \bigcup_{x\in \Sc} B_\epsilon(x)$. Given a
function $s: \Xc \subseteq \real^n \to \real$, $s \in C$ means $s$ is
continuous and $s \in C^n$ means $s$ has a continuous $n^{\text{th}}$
derivative. A function $\alpha: (-a,b) \to \real$ is a class-$\kappa$
function if it is strictly increasing and $\alpha(0) = 0$. The
cardinality, boundary, interior, closure, and convex hull of a set
$\Sc$ are denoted by $|\Sc|$, $\partial \Sc$, $\interior{\Sc}$,
$\closure{\Sc}$, and $\co{\Sc}$, respectively. We write
$F:\Sc \rightrightarrows \Sc'$ to denote that $F$ maps elements of
$\Sc$ to subsets of $\Sc'$.

%%%%%%%%%%%%%%%%%%%%%%%%%%%%%%%%%%%%%%%%%%%%%%%%%%%%%%%%%%%%%%%%%%%%%%%%%%%%%%%%%
% start by specifying the dynamics which will be studied. Consider the
\section{Problem Statement}\label{sec:problem}

Our goal is to construct computationally-inexpensive controllers that
render solutions of possibly discontinuous dynamics forward invariant
with respect to possibly non-smooth sets.
% Before stating the problem mathematically, we first review some
% preliminaries.
We consider a system with control-affine dynamics:
\begin{equation}\label{eq:affSys}
  \dot x = f(x) + G(x)u, 
\end{equation}
with $x \in \Xc \subseteq \real^n$ and $u \in \Uc$, where $\Uc$ is a
convex set in $\real^m$.  We assume that $f:\Xc \to \real^n$ and
$G:\Xc \to \real^{n\times m}$ are piecewise
continuous~\cite{JC:08-csm-yo}, that is, there is a collection of
open, disjoint sets
$\Xc_1, \dots, \Xc_\omega \subseteq \Xc$ with the closure of their union
covering $\Xc$, and such that $(f,G)$ is continuous on each
$\Xc_j$. For simplicity, we further assume that, for each
$j \in \{1,\dots, \omega\}$, there is a continuous map
$(f_j,G_j):\Xc \to \real^n \times \real^{n \times m}$ such that
$(f(x),G(x)) = (f_j(x),G_j(x))$ for all $x \in \Xc_j$.
%
% \marginJC{Why $r_1$ and not simply $r$? As defined, this could be very
% wild, like $(f_1,G_1)$ on the rationals and $(f_2,G_2)$ on the
% irrationals. Like that, I don't think $(f,G)$ would be Lipschitz a.e.}
%
% This dynamics is slightly more general than the
% piecewise continuous dynamics described in~\cite{JC:08-csm-yo}, for
% the set at which $(f,G) = (f_j,G_j)$ is not assumed open.
% %
% \marginJC{Maybe have a dedicated remark where we comment in the
%   nuances of the def, but not in the main text. }
%
We denote the set of active dynamics by
$\Jc(x) \coloneqq \{j \in \{1, \dots, \omega\}\; | \; x \in
\closure{\Xc_{j}}\}$. Accordingly, $\Jc(x)$ is a singleton almost
everywhere.
% {\color{red} At any point in the boundary between different
% $\Xc_i$'s, we take $(f(x),G(x)) = (f_j(x),G_j(x))$ for some
% $j \in \Jc(x)$.}
% %
% \marginJC{Not sure what we want to convey here. You mean that there
%   are potentially many different choices, and we just take one? Is
%   the specific choice relevant? If it's not, then maybe there is no
%   need to say it?}
%

Our assumptions on the system dynamics are not as restrictive as
assuming the Lipschitzness of the dynamics, and not as permissive as
differential inclusions~\cite{JC:08-csm-yo}.  Unlike Lipschitz
dynamics, our model can capture systems with Coulomb friction and
multi-agent systems with discontinuous coordination algorithms. Our
model, however, is more structured than a differential inclusion. This
additional structure allows for convenient control techniques, as it
will become clear in the paper.

% is complex to capture important systems that are not captured by
% mere Lipschitz dynamics such as systems with Coulomb friction and
% multi-agent systems with control algorithms such as
% move-away-from-nearest-neighbor~\cite{JC:08-csm-yo}. Our system
% dynamics, however, are more structured than differential inclusions
% which allows for convenient control techniques as we will show
% throughout the paper.
%
% \marginJC{It's not immediately clear, and we don't articulate it, what
% we win by this more complex definition of the dynamics and of the CBF.}
%
%\marginJC{How about we say that we often refer to $h_i$ as component function later}
%

We next describe the types of set that we want to keep forward
invariant under the dynamics. Let $h_i:\Xc \to \real$,
$i \in \{1,\dots,r\}$ be $C^1$ functions on $\Xc$ with associated
$0$-superlevel sets
$\Cc_i \coloneqq \{x \in \real^n \;|\; h_i(x) \geq 0\} \subseteq
\Xc$. Let
\begin{equation}\label{eq:h}
  h:\Xc \to \real 
\end{equation}
be a piecewise differentiable Lipschitz function. That is, $h$ is
Lipschitz and there are $r$ open, disjoint sets
$\Xc^{h_1}, \dots, \Xc^{h_r}$ with the closure of their union covering
$\Xc$ and $h(x) = h_i(x)$ for all $x \in \Xc^{h_i}$.  We refer to the
functions $\{h_i\}_{i=1}^r$ as the components of $h$, and denote the
set of active components by
$\Ic(x) \coloneqq \{i \in \{1, \dots, r\}\; | \; x \in
\closure{\Xc^{h_i}}\}$. Accordingly, $\Ic(x)$ is a singleton almost
everywhere.  We denote by
$\Cc \coloneqq \{x \in \real^n \;|\; h(x) \geq 0\}$ the $0$-superlevel
set of~$h$.
%
% \marginJC{So we are not saying $h$ is the max or the min of $h_i$'s,
%   right? }
%   \marginMA{yes}
% %

Our aim is to construct a controller $u = k(x)$ for the
dynamics~\eqref{eq:affSys} that renders $\Cc$ forward invariant under
the closed-loop solutions of~\eqref{eq:affSys}. We refer to such
a controller as \emph{safe}.  To formalize this objective, it is
necessary to specify the notion of solution to such discontinuous
dynamics with a possibly discontinuous controller. To do so, we adopt
the widely-used notion of Filippov solution \cite{AFF:13}.

Given a differential equation $\dot x = X(x)$, where
$X: \Xc \subseteq \real^n \to \real^n$, the Filippov set-valued map
for $x \in \Xc$ is
\begin{equation}\label{eq:FilSetMap}
  F[X](x)\coloneqq \bigcap_{\delta > 0} \bigcap_{\eta(\Sc) = 0} \bar{\text{co}}
  \{X(B_\delta(x)\setminus \Sc)\},  
\end{equation}
where $\bar{\text{co}}$ denotes the convex closure and $\eta(\Sc)$ the
Lebesgue measure of $\Sc$. A \emph{Filippov solution of} $\dot x = X(x)$ is an absolutely continuous map $\phi: [t_0,t_1] \to \Xc$ that satisfies
$\dot \phi(t) \in F[X](\phi(t))$ for almost all $t \in [t_0,t_1]$. For
the piecewise-continuous control-affine dynamics with a piecewise
continuous controller $k: \Xc \to \Uc$, the Filippov set-valued map
simplifies to~\cite{JC:08-csm-yo}:
\begin{align}
  &F[f+Gk](x) =\nonumber
  \\
  & \; \co{\{\lim_{\mu \to \infty} \left(f(x_\mu)+G(x_\mu)k(x_\mu)\right)
    \mid x_\mu \to x, x_\mu \notin S\}}, \label{eq:Fil4Piece} 
\end{align}
where $S$ is the set in which $f+Gk$ is discontinuous.
We are now ready to define what we mean by a safe
controller.% which will be instrumental in our problem definition.

\begin{definition}\label{def:controlledInvariance}
  \rm{\textbf{(Safe Feedback Controller and Set Invariance).}
    % Let $\Sc \subseteq \Xc$ be open.
    % %
    % \marginJC{I don't understand the role of $\Sc$. It seems to be an
    %   intermediary between $C$ and $\Xc$. Why do we need it?}
    % %
    A feedback controller $k: \Xc \to \Uc$ for the general
    non-autonomous dynamics $\dot x = \bar f(x,u)$
    %
    % \marginJC{Here and in what follows, you need to rethink this use
    %   of notation again: $f'$ looks like ``derivative of $f$'', and
    %   that's not what we want. }
    % 
    is \emph{safe
      with respect to} $\Cc \subseteq \Xc$ if
  \begin{enumerate}[(i)]
  \item there exists at least one Filippov solution $\phi$ to
    $\dot x = \bar f(x,k(x))$ starting from any point in $\Cc$ and,
  \item all Filippov solutions $\phi$ starting at $\Cc$ remain in
    $\Cc$ for all $t > 0$. \oprocend
  \end{enumerate}
}
\end{definition}

The set $\Cc$ is \emph{(forward) invariant under} $k$ or just
\emph{controlled-invariant} if such a $k$ exists. We use the term
\emph{safeness} when attributed to the control and the term
\emph{safety} when attributed to the set. We state next precisely the
problem tackled in this work.

\begin{problem}\label{prob:getController}
  \rm{Find a controller for~\eqref{eq:affSys} that
    \begin{enumerate}[(i)]
    \item is continuous wherever the dynamics are continuous,
    \item is safe with respect to $\Cc$, and
    \item solves a feasible convex optimization
      problem. \oprocend
    \end{enumerate}}
\end{problem}

By feasible optimization, we mean that the constraints of the program
can always be satisfied by some control values. The requirement that
the controller solves a convex optimization problem is motivated by
the prevalence of QP controllers for the simpler case of Lipschitz
dynamics with smooth safe sets,
cf.~\cite{ADA-SC-ME-GN-KS-PT:19,ADA-XX-JWG-PT:17}. The light
computational effort required to solve such programs
%
% \marginJC{But we say ``an optmization problem'', not a QP. I can write
% optimization problems that are superhard to solve. Maybe you want to
% specify convex optimization problem (at least, those are now
% universally recognized as ``easy to solve'').}
% %
makes them useful for fast online control computation.

We aim here to enforce safety for the more general case of
discontinuous dynamics and non-smooth sets without adding
computational complexity to the control synthesis. Also, as seen from
the first requirement, we only require our controller to be continuous
when the dynamics are. In other words, we do not demand from the
controller regularity properties that the dynamics does not
have. This, we believe, is a reasonable middle ground between
demanding continuity of the controller for discontinuous
dynamics \cite{AI-MG-RGS-WED:24} and allowing discontinuity of the
controller with continuous
dynamics \cite{JU-KG-DP:20,AW-DVD:22,AZ-HGT:22}.

\section{Sufficient Conditions for Safe Control}

We review here the state-of-the-art conditions on any control that
renders a nonsmooth control-affine dynamics of the
form~\eqref{eq:affSys} safe with respect to a given set.
%
% \marginJC{I don't think the reader has enough context to understand
%   what we mean by this: conditions on what? The dynamics? The safety
%   set? The control design?}
%
%This will help us later in proving the safeness of the controller we propose. 
% 
We employ the notion of generalized gradient \cite{FHC:75} of a
Lipschitz function $h$,
\begin{equation*}
  \partial h(x) = \co{ \{\lim_{i \to \infty}{\nabla h(x_i)\; | \; x_i
      \to x, x_i \notin \Omega_h} \}}, 
\end{equation*}
%
%
%\marginJC{Should be $\Omega_h$, no?}
%
where $\Omega_h$ is the zero-measure set on which $h$ is
non-differentiable. For $h$ that satisfies our description in Section~\ref{sec:problem}, we have
\begin{equation}
  \partial h(x) = \co{ \{\nabla h_i(x)\; |\; i \in \Ic(x)\} }. \label{eq:dh}
\end{equation}
We also use the notion of generalized Lie derivative
\cite{JC:08-csm-yo} of a Lipschitz function $h$ with respect to a
set-valued map $F:\Sc \rightrightarrows \real^n$,
% \NA{I forgot if I suggested the powerset notation but it may be
% better to write $F:\Sc \rightrightarrows \real^n$.}
%
\begin{equation*}
  \tilde \Lc_{F}h(x) \coloneqq \{a \in \real \;|\;\exists v \in F(x),
  \forall \zeta \in \partial h(x), a = v^\top\zeta\}.
\end{equation*}
The following result, adapted from~\cite[Thm. 3]{PG-JC-ME:17-csl},
gives a general condition for safeness of a feedback controller.

\begin{theorem}\label{thm:safetyCond} {\rm \textbf{(Sufficient Condition
      for Safe Control).}}
  % Noting that $\Cc$ is the superlevel set of the function $h$
  % described in~\eqref{eq:h},
  % \NA{We should state what $h$ is to make the theorem
  % self-contained. It may be good to define $h$ in an equation, like
  % (1), so we can refer to it here and in several places later on,
  % such as Assumption 1, Lemma 4.1, etc.}
  Given $\epsilon > 0$, let $k:B_\epsilon(\Cc) \to \Uc$ be a feedback
  controller for the non-autonomous dynamics $\dot x = \bar f(x,u)$ and
  let $F$ be the Filippov set-valued map associated to the closed-loop
  dynamics.  Assume that $\dot{x} = \bar f(x,k(x))$ has a Filippov
  solution starting from every point in $\Cc$.
  %
  % \marginJC{We've already introduced (1), $\dot x = X(x)$, and now
  % $\dot x = X(x,u)$. Do we need that many?}
  %
  % \marginMA{We actually do! (1) is the dynamics we work with. $X(x)$
  % is for defining what a solution is. $X(x,u)$ is to define what a
  % safe control is. To avoid confusion, I changed $X(x,u)$ to
  % $f'(x,u)$. What you think?}
  If there exists a class-$\kappa$ function $\alpha$ such that
  \begin{equation}\label{eq:safetyCond}
    \inf \tilde{\Lc}_{F[\bar f(\cdot,k(\cdot))]}h(x) \geq -\alpha(h(x)) ,
  \end{equation}
  in a neighborhood of $\partial \Cc$, then $k$ is safe with respect
  to $\Cc$. More precisely, if~\eqref{eq:safetyCond} is satisfied in a
  neighborhood of $\bar x\in \partial \Cc$, then no Filippov solution
  of $\dot x = \bar f(x,k(x))$ can leave $\Cc$ from $\bar x$.
  %Moreover, any solution starting from the neighborhood where the aforementioned inequality holds converges to $\Cc$. 
\end{theorem}
\begin{pf}
  % \textit{\underline{Proof}:}
  Let $\phi:[t_0,t_1] \to B_\epsilon(\Cc)$ be a Filippov solution of
  $\dot x = \bar f(x,k(x))$ and $\phi(t_0) \in \Cc$. Suppose by
  contradiction that $\phi$ leaves~$\Cc$. Thus, without loss of
  generality, $\phi(t_1) \notin \Cc$, that is, $h(\phi(t_1)) < 0$. By
  the absolute continuity of $h$ and $\phi$ and the facts that
  $h(\phi (t_0)) \geq 0$ and $h(\phi (t_1)) < 0$, there exists
  $\bar t \in [t_0,t_1]$ for which $h(\phi(\bar t)) = 0$ and
  $h(\phi (t)) < 0$ for all $t \in  (\bar t,t_1]$.
  %
  % \marginJC{You mean for $t \in (\bar t,t_1]$? I guess you need to
  % say
  % without loss of generality, b/c in principle the sol could come
  % back in. Here, we're reasoning with the last time it exited and
  % never came back.}
  % \marginMA{Yes this is what I am speaking about but I still don't
  % understand why I should put WLG. I'm just saying by continuity and
  % $h(\phi(t_0))>0$ and $h(\phi(t_1))<0$ there is such a $\bar t$
  % which is the time at which it exited last.}
% 
  To derive a contradiction,
  we use the fact that in a neighborhood of $\phi(\bar t)$, we have
  $\inf {\tilde \Lc_{F}h(x)} \geq -\alpha(h(x))$ and assume without loss of generality that $\phi(t_1)$ is in that neighborhood. Due to absolute
  continuity, the time derivative $\dt{h(\phi(t))}$ exists almost
  everywhere in~$[t_0,t_1]$ and
  $\dt{h(\phi(t))} \in \tilde \Lc_{F}h(\phi(t))$ for almost every
  $t \in [t_0,t_1]$~\cite[Lem. 1]{AB-FC:99}. Thus,
  $\dt{h(\phi (t))} \geq -\alpha(h(t))$ almost everywhere in the
  interval~$[\bar t,t_1]$. This gives
  \begin{align*}
    0&> h(\phi(t_1)) - h(\phi (\bar t)) = \int_{\bar t}^{
       t_1}\dt{h(\phi (t))}dt
    \\ 
     & \geq -\int_{\bar t}^{t_1}\alpha(h(\phi (t)))dt.
  \end{align*}
  Recalling that $\alpha(h(\phi (t))) < 0$ on $(\bar t, t_1)$, as $\alpha$ is a class-$\kappa$ function, the
  last inequality leads to a contradiction.  \oprocend
\end{pf}

Theorem~\ref{thm:safetyCond} is a slight variation
of \cite[Thm. 3]{PG-JC-ME:17-csl}: the difference is that
condition \eqref{eq:safetyCond} is required to hold on a neighborhood
of $\partial \Cc$, whereas in \cite[Thm. 3]{PG-JC-ME:17-csl} it is
required to hold on a neighborhood of~$\Cc$. The formulation here
touches on whether a trajectory can leave the set from a specific
single point, motivated by our ensuing discussion. Condition \eqref{eq:safetyCond} is less strict than the requirement in \cite{MG-AI-RGS-WED:22f} that $\zeta^\top v \geq \alpha(h(x))$ for all $\zeta \in F[\bar f(x,k(x))]$ and all $v \in \partial h(x)$. Note also that the safety
condition \eqref{eq:safetyCond} does not require continuity of $k$.
From these two perspectives, Theorem~\ref{thm:safetyCond} is a
generalization of the safety conditions in \cite{MG-AI-RGS-WED:22f}.

% \NA{Do you mean \cite{MG-AI-RGS-WED:22f} or \cite{PG-JC-ME:17-csl}?
% The comparisons until now were with respect to
% \cite{PG-JC-ME:17-csl}.}.
% We are now ready to address Problem \ref{prob:getController}.
% 
% \marginJC{I don't fully understand the logic of presentation for this
%   section. Why don't we just address problem 1 after having stated it
%   in the previous section? I guess what we're missing is a discussion
%   (probably at the beginning of this section) making it explicit that
%   this discussion here provides guidance as to what properties to
%   enforce with our controller design, no? And here we can say that
%   having established it, we proceed to the design.}
%\marginMA{I think the discussion now is clearer. What u think?}
% %

\section{Safe Control of Piece-wise Continuous Control-Affine Dynamics}

Having described a sufficient condition for a feedback controller to
be safe, here we tackle the problem of actually synthesizing the
controller. To do so, our approach takes inspiration from studying the
safety limitations of controllers developed for continuous
settings. We start by considering a controller that we term
\emph{active-component QP controller}. It is a direct extension of the
controller used in~\cite{LL-DVD:20,AW-DVD:22,AZ-HGT:22} to verify
safety with non-smooth safe sets but with continuous dynamics. We
study the extent to which this controller can yield safety with
discontinuities in the dynamics.  The naming of the active-component
QP controller emphasizes the fact that its safety constraints at any
state $x$ are only concerned with the components of $h$ active at that
particular $x$. We show that this controller actually fails to enforce
safety at points of non-differentiability of $h$. This motivates our
ensuing control design to overcome these limitations, termed the
\emph{all-components QP controller}, and establish its safeness,
feasibility, and regularity properties.
% We finally use the feasibility result to improve the all-components
% QP controller into the \emph{all-components adaptive QP controller}
% which can readily be applied in practice.

\subsection{Active-Component QP Controller}
The widely-used QP controller~\cite{ADA-XX-JWG-PT:17} for a
differentiable CBF~$h$ solves a quadratic program, with the constraint
being the classical CBF condition
\begin{equation}\label{eq:cbfCond}
  \nabla h(x)^\top (f(x) + G(x)u) + \alpha(h(x)) \geq 0,
\end{equation}
where $\alpha$ is a class-$\kappa$ function. The constraint is
guaranteed to be feasible because of the definition of CBF.
%
% \marginJC{I think the placement of this assumption exactly here is a
%   bit distracting -- I'm writing this margin now *after* having
%   written the ones that follow. The parallelism between (6) and the
%   constraints in (7) is clear, but not so much with the assumption. So
% I think it's better to move it to before Th 4.2, rather than in the
% middle of getting from (6) to (7).}
%
% \marginJC{Something I don't like is that, if (1) happened to be
%   continuous and the CBF differentiable, then this assumption does not
%   recover the classical CBF condition. Why only on $\partial C$? Why
%   no $\alpha(h(x))$?}
% %
% %
%   \marginJC{But this is not the classical CBF condition, which
%   allows for tangency. And also, the classical CBF condition has the
%   ``graceful violation'' part in the interior of $C$, which here we
%   do not comment on.}
% \marginMA{I included a discussion of why we
%   don't allow the tangency. But the graceful violation thing I
%   didn't discuss since if CBF condition is satisfied on boundary
%   then there is always $\alpha$ that make it work for the
%   interior. Also this grace violation is only introduced to make the
%   control definition more convenient. I feel like a full discussion
%   here might be unnecessarily long.}
%
The active-component QP controller is the analogue of this classical
QP controller, and is given by the following QP:
\begin{align}
  \uact(x) & \coloneqq \argmin_{u \in \Uc} {u^\top Q(x)u +
             b(x)^\top u} \label{eq:qp1}
  \\ 
  \text{s.t. } &\nabla h_i(x)^\top (f(x) + G(x)u) + \alpha(h_i(x))
                 \geq 0,\; \forall i \in \Ic(x) \nonumber 
  %& \hskip 2cm \forall i \in \Ic(x), \nonumber
\end{align}
where $Q:\Xc \to \real^{m \times m}$ and $b:\Xc \to \real^{m}$ are
Lipschitz, $Q(x)$ is positive definite on $\Xc$, and $\alpha$ is a
class-$\kappa$ function. This controller can be seen as minimally
constrained, in that it only asks that the safety constraint be
satisfied for the active component of $h$, that is, $h_i$ with
$i \in \Ic(x)$, and for the dynamics at $x$ only, hence the name
\emph{active-component QP controller}. A similar controller is
introduced in works~\cite{LL-DVD:20,AW-DVD:22,AZ-HGT:22} involving STL
specifications for settings with continuous dynamics.

Similarly to what is done for the classical QP controller, we need to
make an assumption on the function $h$ to ensure the constraints in
the active-component QP controller are feasible.

\begin{assumption}\label{as:genCBFcond}
  \rm{\textbf{(Non-smooth Version of CBF Condition).} Given the system
    dynamics~\eqref{eq:affSys} with piecewise continuous structure and
    a piecewise differentiable function $h$, let
    $\tilde \Ic:\Xc \rightrightarrows \{1,\dots,r\}$ be a set-valued
    map such that $\Ic(x) \subseteq \tilde \Ic(x)$ for all
    $x \in \Xc$.  We assume that for all $x \in \partial \Cc$, there
    exists $u_x \in \Uc$ such that
  \begin{align*}
    \nabla h_i(x)^\top (f_j(x) + G_j(x)u_x) > 0 ,
  \end{align*}
  for all $i \in \tilde \Ic(x)$ and $j \in \Jc(x)$.}\oprocend
\end{assumption}

Assumption~\ref{as:genCBFcond} simply requires that there are control
values that steer the system from the boundary of $\Cc$ to its
interior. It can be shown that the satisfaction of
Assumption~\ref{as:genCBFcond} with $\tilde \Ic(x) = \Ic(x)$ is all
that is needed for the existence of a safe
controller~\cite[Thm. 1]{MG-AI-RGS-WED:22f}. However, for our purpose
here, which is deriving an explicit QP controller, we will need the
satisfaction of Assumption~\ref{as:genCBFcond} with $\tilde \Ic(x)$
that is slightly larger than $\Ic(x)$ at a few states -- this will be
made explicit in our feasibility analysis in
Theorem~\ref{thm:u2feasible}. The classical CBF
definition~\cite{ADA-SC-ME-GN-KS-PT:19} is less strict than
Assumption~\ref{as:genCBFcond} in that it only asks for the existence
of a control value that steers the dynamics to the interior of the
safe set or tangentially to it. The stricter requirement here is
intended to remove difficulties related to regularity and boundedness
of the optimization-based controller \cite{MA-NA-JC:23-tac} that are
allowed for by the classical CBF definition.
  %
% \marginJC{This sounds a bit mysterious. What difficulties? I
%   understand we don't want to meander, but maybe something like
%   ``difficulties related to bla, bla'' to help the reader. Bla, bla
%   could be ``boundedness and regularity of the resulting
%   controller'' or something like that.}
% 
Note also that the constraint in \eqref{eq:qp1} is less restrictive
than the condition in Assumption~\ref{as:genCBFcond}. The latter
requires the safety condition to be satisfied for all $j \in \Jc(x)$,
while the former requires it only for the dynamics at the state at
which the controller is evaluated. As we show below, the reason for
assuming more than the mere feasibility of the constraint
of~\eqref{eq:qp1} is to attain regularity properties for the
controller $\uact$.

We now study the extent to which the proposed active-component QP
controller ensures safety in discontinuous settings.
% \NA{This makes it sound like someone else had already proposed this
% controller. Is this true? What is different in our formulation?}.
The abrupt changes in the functions defining the constraints
in \eqref{eq:qp1} generally introduce discontinuities
in $\uact$. Despite this, the following result characterizes its
regularity properties under Assumption \ref{as:genCBFcond}.

\begin{theorem}\label{thm:discont} {\rm \textbf{(Regularity Properties
      of Active-Components QP controller).}}
  If Assumption~\ref{as:genCBFcond} holds with
  $\tilde \Ic(x) = \Ic(x)$, then in a neighborhood
  $\Nc_{\partial \Cc}$ of $\partial \Cc$, the active-components
  controller $\uact$ defined by \eqref{eq:qp1} with any class-$\kappa$
  function $\alpha$ is
  \begin{enumerate}[(i)]
  \item feasible,
    \item single-valued,
    \item continuous almost everywhere,
    \item bounded in a bounded neighborhood of every point
      in~$\Nc_{\partial \Cc}$, and
    \item yields a Filippov solution to~\eqref{eq:affSys} with
      $u = \uact(x)$ from every point in $\Nc_{\partial \Cc}$.
  \end{enumerate}  
\end{theorem}
\begin{pf}
  We first choose the neighborhood $\Nc_{\partial \Cc}$. For all
  $x \in \partial \Cc$ and all $i \in \Ic(x)$ and $j \in \Jc(x)$,
  $\nabla h_i(y)^\top(f_j(y)+G_j(y)u_x)$
  % $\NA{Do you mean the function is continuous? Perhaps, there should
  % be no ``$>0$".}
  is continuous in $y$ in a neighborhood of $x$.
  % %
  % \marginJC{Why? There is nothing in Assumption 1 that says that $u_x$
  %   changes continuously with $x$, no?}
  % % 
  % where
  % $u_x$ of Assumption~\ref{as:genCBFcond} is considered constant.
  %
  % \marginJC{"...is considered constant"? I don't know what this
  % means. Who is considering is constant and why is this reasonable?}
  % \marginMA{I mean the function of $x$ and
  % $u_x$ is continuous in
  % $x$. I think it is clearer now without my sloppy addition.}
  %
  By Assumption~\ref{as:genCBFcond}, keeping in mind the continuity of
  $\alpha$ and that $\alpha(h_i(x)) = 0$ for all $i \in \Ic(x)$ at $x
  \in \partial \Cc$, there is a neighborhood $\Nc_x^{i,j}$ of
  $x$ such that $\nabla h_i(y)(f_j(y)+G_j(y)u_x) +
  \alpha(h_i(y))>0$ for every $y \in
  \Nc_x^{i,j}$. We define then $\Nc_{\partial \Cc} = \cup_{x \in
    \partial \Cc} (\Mc_x \cap (\cap_{i,j} \Nc_x^{i,j}) ) \cap \Xc$,
  % 
  % \marginJC{How about $(\cup_{x \in \partial \Cc} \Mc_x \cap
  % \cap_{i,j} \Nc_x^{i,j} ) \cap \Xc$ instead?}
  % 
  where
  $\Mc_x$ is given by Lemma~\ref{lem:ij}. We now prove the statements
  of the result:
  \begin{enumerate}[(i)]
  \item Noting that $(f,G) = (f_j,G_j)$ for some $j \in
    \Jc(x)$, the constraint in~\eqref{eq:qp1} is always satisfied by
    some $u_x$ for any point $x$ in $ \Nc_{\partial
      \Cc}$, proving feasibility.
  
  \item For every $x \in \Nc_{\partial
      \Cc}$, the program~\eqref{eq:qp1} has a strictly convex
    objective and a nonempty closed constraint set. Thus the function
    that equals the objective where the constraints are satisfied and
    equals
    $\infty$ otherwise satisfies the premises
    of \cite[Thm.~1.9]{RTR-RJBW:98}. Thus, the set of minimizers is
    not empty.  By \cite[Thm.~2.6]{RTR-RJBW:98}
    %
    % \marginJC{What does 2.6 refer to?  Is it a theorem, or a chapter,
    %   or what?}
    %
    the minimizer is unique.

  \item Since $\Ic(x)$ and $\Jc(x)$ are singletons almost everywhere,
    for almost every $x \in \Nc_{\partial \Cc}$, there is a
    neighborhood $\Nc_x$ of $x$ with
    $\Nc_x \subseteq \Nc_{\partial \Cc}$ such that, for all $y$ in
    $\Nc_x$, $(h(y),\nabla h(y)) = (h_i(y),\nabla h_i(y))$ (where $i$
    is the only element in $\Ic(x)$) and
    $(f(y),G(y)) = (f_j(y),G_j(y))$ (where $j$ is the only element in
    $\Jc(x)$). Thus, $\nabla h(x)^\top(f(x) + G(x)u) + \alpha(h(x))$
    is continuous in both $x$ and $u$ and convex in $u$ for almost
    every~$x$. Keeping in mind the convexity of $\Uc$ and the
    existence of $u_x$ that strictly satisfies the constraint, the
    set-valued map
    $\Nc_{\partial \Cc} \ni x \mapsto \{u \;|\; u \; \text{satisfies
      \eqref{eq:qp1} at $x$}\} \subset \Uc$, is continuous in a
    neighborhood of $x$
    %
    %\marginJC{``continuous around $x$'', you mean ``continuous for
     % almost every $x$''?}
    %
    by~\cite[Thms. 10 \& 12]{WWH:73}.
    %
    %\marginJC{Is that the right numbering for the theorems in that
     % paper?}
     %\marginMA{Yup! The author there uses 'open' and 'close' for lsc and usc.}
    %
    Thus for almost every
    $x \in \Nc_{\partial \Cc}$, $\uact$ is continuous at $x$
    by~\cite[Cor. 8.1]{WWH:73}.

  \item By definition of $\Nc_{\partial \Cc}$, for all
    $\bar x \in \Nc_{\partial \Cc}$ there exists $x \in \partial \Cc$
    such that
    $\nabla h_i(\bar x)^\top(f_j(\bar x) + G_j(\bar x)u_x) +
    \alpha(h_i(\bar x)) > 0$ for all $i \in \Ic(\bar x)$ and
    $j \in \Jc(\bar x)$. By continuity of $\nabla h_i$, $f_i$, and
    $G_i$ there is a neighborhood $\Nc$ of $\bar x$ such that
    $\nabla h_i(y)^\top(f_j(y) + G_j(y)u_x) + \alpha(h_i(y)) > 0$ for
    all $y \in \Nc$ for all $i \in \Ic(\bar x)$ and
    $j \in \Jc(\bar x)$. By Lemma~\ref{lem:ij}, the neighborhood $\Nc$
    can be chosen such that the strict inequality holds for all
    $i \in \Ic(y)$ and $j \in \Jc(y)$. Hence, $u_{\bar x}$ is feasible
    in program~\eqref{eq:qp1} for all $y \in \Nc$, which implies
    %\NA{Should we have $Q$ instead of $H$ below?}
    \begin{align*}
      \uact(y)^\top Q(y)\uact(y)&+b(y)^\top \uact(y) \\
      \leq &u_{\bar x}^\top Q(y)u_{\bar x} +b(y)^\top u_{\bar x}.
    \end{align*}
    Thus, invoking the Lipschitzness of $Q$
    %
    %\marginJC{$H$??}
    %
    and $b$, and the positive definiteness of $Q(\cdot)$, we deduce
    that $\|\uact\|$ is bounded on a neighborhood of $\bar x$.
    %
    % \marginJC{I'm not following the argument here: what role does
    % Lipschitzness play to show boundedness?}
    % \marginMA{Lipschitzness implies boundedness in any bounded
    % neighborhood, here $\Nc$. So $Q(y)$ and $b(y)$ don't blow up and
    % so the right side of the inequality is finite and hence the left
    % side is.}
    %
    
  \item By the assumed piecewise continuous structure of the dynamics
    \eqref{eq:affSys} and the fact that $\uact$ is continuous almost
    everywhere and bounded on a neighborhood of every point, this
    statement is immediate
    from~\cite[Prop. 1]{JC:08-csm-yo}. \oprocend
  \end{enumerate}
\end{pf}

% The assumption in Theorem~\ref{thm:discont} that $\Jc(x)$
% is a singleton means that that there is only one active dynamics. This is satisfied almost everywhere unless two
% dynamics or components of $h$ are equal in a set of positive
% measure. But such violation can always be countered by changing the
% value of one of those components in the set of overlap while
% preserving the assumed regularity, that is, smoothness of $h_i$ or
% continuity for $f_i$ and~$G_i$.
%
% \marginJC{Changing how? And this change should be made so that
%   smoothness is preserved -- so you mean ``changing... so that
%   smoothness is preserved''?}
%
%
% \marginJC{We do not comment on the hypothesis ``Assume $\Ic(x)$ and
%   $\Jc(x)$ are singletons almost everywhere in $\Xc$''}
% \marginJC{With the re-placement of the assumption right before Th 4.1,
%   this red paragraph could be right after the assumption, preceding
%   the theorem.}

Next, we study the safeness properties of the active component QP
controller $\uact$.  The following result shows that trajectories
of~\eqref{eq:affSys} under $\uact$ can be kept from leaving the set
from the points at which $h$ is differentiable.
% which is almost everywhere given the assumed structure of~$h$.

\begin{theorem}\label{thm:unsafeNondiff} {\rm \textbf{(Safeness of
      Active-Components QP Controller at Points of Smoothness).}}
  Under the assumptions of Theorem~\ref{thm:discont}, let
  $k: \Cc \cup \Nc_{\partial \Cc} \to \Uc$ be a controller that gives
  a solution from any initial condition in $\Cc$ and is such that
  $k(x) = \uact(x)$ for all $x \in \Nc_{\partial \Cc}$. Let
  $\phi:[t_0,t_1] \to \Nc_{\partial \Cc}$ be a Filippov solution
  of~\eqref{eq:affSys} under $u = k(x)$ with $\phi(t_0) \in \Cc$. If
  $\phi (t) \notin \Cc$, for some $t>t_0$ then there exists
  $\bar t \in [t_0,t)$ such that $h(\phi(\bar t)) = 0$ and
  $\Ic(\phi(\bar t))$ is not a singleton.
  %
  % \marginJC{I'd rather state this for any $t$ such that $\phi (t)
  %   \notin \Cc$. The fact that you only state it for the final time of
  % the trajectory might make the reader interpret that if this happens
  % before, the result does not hold. }
  %
\end{theorem}
\begin{pf}
  Since $h(\phi (t_0)) \geq 0$ and $h(\phi (t)) < 0$ there exists
  $\bar t \in [t_0,t)$ for which $h(\phi(\bar t)) = 0$ and
  $h(\phi (t)) < 0$ for all $t \in (\bar t,t)$.  We reason by
  contradiction and suppose that $\Ic(\phi(\bar t))$ is the singleton
  $\{i\}$. Therefore, $h$ is differentiable on a neighborhood of
  $\phi (\bar t)$ according to our assumptions on the structure of
  $h$. The gradient of $h$ at $\phi (\bar t)$ is
  $\nabla h_i(\phi(\bar t))$.
  %
  %\marginJC{You mean $\nabla h_i(\phi(\bar t))$??}
  %
  We now prove that
  $\inf {\tilde \Lc_{F[f+Gk]}h(x)} \geq -\alpha(h(x))$ in a
  neighborhood of $\phi(\bar t)$, which by
  Theorem~\ref{thm:safetyCond} implies that $\phi$ does not leave
  $\Cc$ from $\phi(\bar t)$ which contradicts the supposition.
  % 
  % \marginJC{This is poorly expressed. It does not contradict the
  %   theorem. If this inequality holds then, \textbf{by Theorem 3.1},
  %   this implies safety, which is a contradiction with $\phi$ stepping
  %   outside of $C$.}
  % 
  For any $a \in \tilde \Lc_{F[f+Gk]}h(x)$ where $x$ is sufficiently
  close to $\phi (\bar t)$ there is $\zeta \in F[f+Gk](x)$ such that
  $a = \nabla h_i(x)^\top \zeta$. Since $\zeta \in F[f+Gk](x)$,
  by~\eqref{eq:Fil4Piece}, there is a finite number say $p$, of
  sequences $\{\{x_\mu^1\}, \dots, \{x_\mu^p\}\}$
  % %
  % \marginJC{You mean to say that $f+Gk$ is continuous on these points
  %   $x_\mu^j$?}
  % \marginMA{No. I mean the form of any vector $\zeta$ in the Fil. set map is so and so as given in \eqref{eq:Fil4Piece}}
  % %
  such that $x^j_\mu \to x$ as $\mu \to \infty$ and
  $\zeta = \sum_{j=1}^{p}\sigma_j \lim_{\mu \to \infty}(f(x_\mu^j) +
  G(x_\mu^j)k(x_\mu^j))$, where $\sigma_j$'s are the constants of the
  convex combination. Thus,
  %\NA{What is $p$?}
  \begin{align*}
    \nabla h(x)^\top  \zeta
    &= \nabla h_i(x)^\top  \sum_{j=1}^{p}\sigma_j
      \lim_{\mu \to \infty}(f(x_\mu^j) +
      G(x_\mu^j)k(x_\mu^j))
    \\ 
    & =  \sum_{j=1}^{p}\sigma_j\lim_{\mu \to
      \infty} \nabla h_i(x_\mu^j)^\top(f(x_\mu^j) +
      G(x_\mu^j)\uact (x_\mu^j))
    \\
    &\geq - \sum_{j=1}^{p}\sigma_j\lim_{\mu \to
      \infty}\alpha(h_i(x_\mu^j)) = -\alpha(h(x)). 
  \end{align*}
  The equality is due to the continuity of
  $\nabla h_i$ at $x$
  % 
  %\marginJC{You mean the continuity of $\nabla h_i$??}
  % 
  and the last inequality is implied by the constraint
  in~\eqref{eq:qp1} which $\uact$ satisfies. Thus in a neighborhood of
  $\phi(\bar t)$, $\inf {\tilde \Lc_{F[f+Gk]}h(x)}\geq -\alpha(h(x))$.
  %
  % \marginJC{ $\inf {\tilde \Lc_{F}h(x)}\geq -\alpha(h(x))$? Why do we
  %   change the notation from $F[f+Gk]$?}
  %
  %This contradicts Theorem~\ref{thm:safetyCond}.
  \oprocend
  %
  % \marginJC{We need more care in expressing the arguments -- see
  %   margin below.}
  %
\end{pf}

A consequence of the previous result is that the active-component QP
controller is good enough when $h$ is differentiable even if the
dynamics is discontinuous.
% and, consequently, discontinuous active-component QP controller.

\begin{corollary}\label{cor:discSmooth} {\rm
    \textbf{(Active-Components QP Controller Is Safe for Smooth
      Set).}}
  Under the assumptions of Theorem~\ref{thm:discont}, let
  $k: \Cc \cup \Nc_{\partial \Cc} \to \Uc$ be a feedback controller
  defined as in Theorem~\ref{thm:unsafeNondiff}. If $\Ic(x)$ is a
  singleton for all $x \in \partial \Cc$, then $k$ is safe with
  respect to~$\Cc$.
\end{corollary}

Note that Theorem~\ref{thm:unsafeNondiff} leaves open the possibility
of the closed-loop system under $\uact$ being unsafe at points of
nonsmoothness of $h$. The following example shows that this is indeed possible.

% One should not jump from Theorem~\ref{thm:unsafeNondiff} to deduce
% full safety of $\Cc$ under $\uact$ in general. The following example
% shows that the constraint on $\uact$ is not enough to ensure safety of
% a superlevel set of a non-differentiable $h$ for general discontinuous
% dynamics.

% \NA{This is a minor comment but Thm 4.3 is called ``Safeness of
% Active-Component QP Controller". This is followed by an example
% called ``Un-safe Active-Components QP Controller". This may seem
% contradictory to the less careful reader.}

\begin{figure}
  \centering
  \begin{tikzpicture}
    \begin{axis}[
      axis lines = left,
      xlabel = $x_1$,
      ylabel = $x_2$,
      xmin = -1,
      xmax = 2,
      ymin = -2,
      ymax = 2
      ]
      \draw (axis cs:-1,-2) --  (axis cs:1,0);
      \draw (axis cs: -0.5,-1) node {$h_1(x)\geq 0$};
      \draw (axis cs:-1,2) -- (axis cs:1,0);
      \draw (axis cs: -0.5,1) node {$h_2(x)\geq 0$};
      \draw[blue,->](axis cs:1,0)--(axis cs:1.7,0.7);
      \draw[blue,->](axis cs:1,0)--(axis cs:1.7,-0.7);
      \draw[red,->](axis cs:1,0)--node[above right]{$\zeta$}(axis cs:1.85,0);
      \node at (axis cs:1,0)[circle,fill,inner sep=2pt]{};
      \draw[dotted](axis cs:-1,0) -- (axis cs:2,0);
      \node[text=blue] at (axis cs:1.5,.87) {$f+G_1\uact$};
      \node[text=blue] at (axis cs:1.5,-.87) {$f+G_2\uact$};
      \node[] at (axis cs:0,0.25) {$h(x)\geq0$};
      %\draw (axis cs:-1,0) -- (axis cs:2,0);
    \end{axis}
  \end{tikzpicture}
  \caption{Illustration for Example~\ref{ex:unsafeDisc} of the
    closed-loop dynamics under the controller $\uact$ at the corner
    point of the safe set. The controller prevents trajectories from
    violating safety constraints at points of smoothness of $h$,
    cf. Theorem~\ref{thm:unsafeNondiff}, but does not prevent
    violating safety from points of non-smoothness.}
  \label{fig:2infty}
\end{figure}
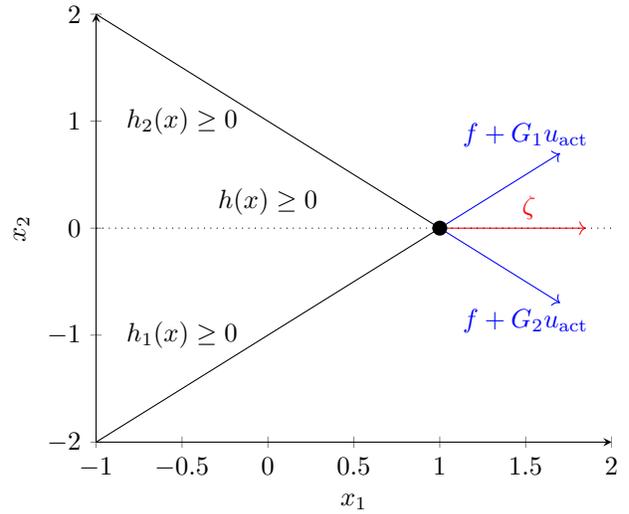

\begin{example}\label{ex:unsafeDisc}
  \rm{\textbf{(Un-safe Active-Components QP Controller).}  On
    $\real^2$, let $f(x) = (1,0)$, $G_1(x) = (-2,1)$,
    $G_2(x) = (-2,-1)$ and consider the discontinuous control-affine
    dynamics defined by
  \begin{equation*}
    (f(x),G(x)) = 
    \begin{cases}
      (f(x),G_1(x)), & x_2 < 0,
      \\
      (f(x),G_2(x)), & x_2 > 0.
    \end{cases}
  \end{equation*}
  Consider the function $h(x)= \min\{h_1(x),h_2(x)\}$, where
  $h_1(x) = x_2-x_1+1$ and $h_2(x) = -x_2-x_1+1$. Note that the
  hypotheses of Theorem~\ref{thm:discont} hold.
  % 
  % \marginJC{By ``premise'' you mean hypotheses? There are more than
  % one, so I'd say the hypotheses of bla hold.}
  % 
  Taking $Q(x)= 1$, $b = 0$, and $\alpha(r) = r$ in~\eqref{eq:qp1}
  gives the following explicit expression of~$\uact$ restricted to the
  region of interest in the state space,
  $\Yc = \{x\;|\; -x_1<x_2<x_1\}$.
  %
  %\marginJC{Notation for SET and use it later too.}
  %
  \begin{equation*}
    {\uact}_{|\Yc}(x) =
    \begin{cases}
      -\frac{1}{3}(x_2 - x_1), & x_2 < 0,
      \\
      \frac{1}{3}(x_2 + x_1), & x_2 > 0.
    \end{cases}
  \end{equation*}
  %
  % %\marginJC{This is an incomplete definition. What happens at all the
  %   other points? You mean to say that $\uact(x) = 0$ otherwise??}
  % \marginMA{Not a definition. An expression for the region of interest only.}
  %
  The closed-loop dynamics is continuous around $x = (1,0)$ except on
  the set with $x_2=0$, and is given in $\Yc$ by
  \begin{equation*}%\label{eq:ExClosedLoop}
    \dot x_{\Yc}=
    \begin{cases}
      (1,0) + (2,-1)\frac{1}{3}(x_2 - x_1), &  x_2 < 0,
      \\
      (1,0) + (-2,-1)\frac{1}{3}(x_2 + x_1), & x_2 > 0.
    \end{cases}
  \end{equation*}
  By~\eqref{eq:Fil4Piece}, in a neighborhood of $(1,0)$, the Filippov
  set-valued map $F[f+G\uact](x)$ is the set of all convex combinations of
  the two cases in this expression. One such vector for points in the neighborhood of $(1,0)$ with $x_2 = 0$ is
  %
  % \marginJC{This is almost impossible to parse through for the
  %   reader. For starters, we're making claims about the closed-loop
  %   dynamics, but we have not written it explicitly for the reader's
  %   convenience. I don't understand what you want to say with ``the
  %   Filippov set-valued map $F[f+G\uact]$ is what determines the
  %   Filippov solutions passing through that set'': of course the
  %   Filippov set-valued map determines what happens with the Filippov
  %   solutions, that's a tautology.}
  % %
  % \marginJC{What is this mysterious $\zeta$ that all of a sudden shows
  %   up here?}
  %
  \begin{align*}
    \zeta
    &= \frac{1}{2}\left (f(x)+\frac{1}{3} G_1(x) x_1 + f(x) + \frac{1}{3}G_2(x)x_1
      \right )
    \\
    &=
      \begin{bmatrix}
        1 - \frac{2}{3}x_1
        \\
        0
      \end{bmatrix}
      \in F[f+G\uact](x).
  \end{align*}
  %
  % \marginJC{Why only the convex combination $(1/2,1/2)$ and not any
  % arbitrary convex combination of both directions. We don't say.}
  %
  However, $\dot x_1 = 1 - \tfrac{2}{3} x_1$, starting from
  $x(0) = (1,0)$, has the solution $x_1(t) = 3/2-1/2e^{-2t/3}$, which
  leaves the $0$-superlevel set of $h$.
  % \NA{Perhaps, saying ``leaves the superlevel set of $h$'' here
  % would be more clear.}
  %
  % \marginJC{The fact that there is 1 vector in the set-valued map that
  %   faces in the wrong direction does not mean safety is violated. One
  %   needs to justify that there exists a Filippov solution that
  %   corresponds to that vector -- there might be none. So, instead of
  %   this argument, it's better to show that sliding along the $x$-axis
  %   is a Filippov solution.}
  %
  Therefore, $\uact$ is not safe with respect to the $0$-superlevel
  set of~$h$. Figure~\ref{fig:2infty} illustrates this.
}\oprocend
\end{example}

%
% \marginJC{In the plot, it does not make sense to write $h_1$ next to a
%   line. $h_1$ is a function! You mean $h(x) = 0$?? Same for
%   $h_2$. Also, better $h(x)>0$. Also, placement of symbols could be
%   farther from lines.}
%

Example~\ref{ex:unsafeDisc} shows that the obvious counterpart of the
popular QP controller is not safe with respect to a non-smooth
set. The reason for this lack of safety, as
Example~\ref{ex:unsafeDisc} shows, is that the controller enforces a
safety constraint only with respect to the active $h_i$ at a specific
state and it does not enforce any constraints with respect to other
$h_{i'}$'s no matter how near the states of their activity are. This
might generate limiting dynamics that are unsafe with respect to
them. % This dynamics then shows itself in the Filippov set-valued map
% and thus allows for unsafe trajectories.

%
% \marginJC{Maybe this last part (starting from here) could be a
%   dedicated remark, since we seem to survey various solution
%   approaches in the literature. Btw, no reference to my work w/ Paul?}
% \marginMA{The controller proposed by your work with Paul imposes constraints on all $h_i$'s. So I referenced it as a precursor to our all-component controller in the first paragraph of the next section.}
% %
\begin{remark}\rm{\textbf{(Methods of Safety Enforcement at Non-Smooth
      Points).}  If a controller is to be safe, it should impose
    safety constraints with respect to the functions $h_{i'}$'s which
    are close enough to being active. This can be done in multiple
    ways. One way, adopted in~\cite{MV-JT:23}, is to enforce the
    constraint of~\eqref{eq:qp1} for all $i \in \tilde \Ic(x)$, where
    $\tilde \Ic(x)$ is chosen to include the indices of all $h_{i'}$'s
    which are close enough to being active. Another way, used
    in~\cite{MG-AI-RGS-WED:22f}, is to enforce a constraint on the
    Filippov set-valued map rather than on individual active dynamics
    and enforce the continuity of the controller. As shown
    in \cite{MG-AI-RGS-WED:22f}, under those two conditions, which are
    guaranteed to be satisfiable under Assumption~\ref{as:genCBFcond},
    safety is attained. Both of those methods require online
    calculations other than those required by a mere QP: in the
    former, to find out $h_{i'}$'s which are close enough to being
    active, and in the latter to calculate the generalized Lie
    derivative for a differential inclusion or a similar
    quantity. Moreover, the first method does not produce a continuous
    controller with continuous dynamics and the synthesis of a
    continuous controller using the second method is not presented explicitly
    in \cite{MG-AI-RGS-WED:22f}. In the following section, we propose
    a safe controller that overcomes these difficulties.} \oprocend
\end{remark}

\subsection{All-Components QP Controller}
In this section, we propose a safe controller that does not require
more online calculation than a standard QP and is continuous when the
dynamics is continous. We further prove its existence for all safe
sets described as nested unions and intersections of the superlevel
sets of the components $h_i$'s. The controller is obtained from a
program of the following form
% \NA{Some transposes are missing for $\nabla h_i(x)$ below and
% onwards.}
\begin{align}
  \uall(x)
  & \coloneqq \argmin_{u \in \Uc} {u^\top Q(x)u +
    b(x)^\top u} \label{eq:qp2}
  \\
  \text{s.t. } &\nabla h_i(x)^\top (f(x) + G(x)u) + \alpha(h_i(x))
                 +\beta_i(H(x)) \geq 0, \nonumber
  \\
  & \hskip 2cm \forall i \in \{1,\dots,r\}, \nonumber
\end{align}
with the same conditions on $Q$, $b$, and $\alpha$ as those
in~\eqref{eq:qp1}, where $H(x) \coloneqq (h_1(x), \dots, h_r(x))$ and
$\beta_i:\Xc \to \real$
%
% \marginJC{Shouldn't $\beta_i$ be restricted to nonnegative (or
%   nonpositive, I cannot remember) values? }
%   \marginMA{No need really.}
% \marginJC{But later, when we say, ``the presence of the constraints
%   facilitates the satisfaction of the constraints'', that wouldn't be
%   true if $\beta_i$ is negative -- in that case, it makes it more
%   difficult, no?}
%   \marginMA{True. I reworded that later part accordingly.}
%
any continuous function with the property that $\beta_i(H(x)) = 0$
when $i \in \Ic(x)$.  We provide an exact form of $\beta_i$ later when
establishing the feasibility of~\eqref{eq:qp2}. Notice that,
unlike~\eqref{eq:qp1}, the constraints in~\eqref{eq:qp2} are enforced
for all $i$, and thus the terminology \emph{all-components QP
  controller}. The idea of enforcing safety constraints on all
components $h_i$ appears also in~\cite{PG-JC-ME:17-csl}. There,
however, the structure of the safe set is limited to an intersection
of the superlevel sets of the components $h_i$'s and the feasibility
of the program is not studied. Note also that, different
from~\eqref{eq:qp1} and the controller in~\cite{PG-JC-ME:17-csl}, the
functions $\beta_i$ appear in the constraints of~\eqref{eq:qp2}.
We refer to these functions as \emph{transition functions} and 
will choose them later in a manner that facilitates the
satisfaction of the constraints for the inactive~$h_i$'s.
%
% \marginJC{Feel free to change the name -- if you do, propagate it to
%   later too.}
%
% We provide a detailed interpretation of Program~\eqref{eq:qp2} in
% Remark~\ref{rem:constraintsMeaning}.  The constraints
% in~\eqref{eq:qp2} are generally more restrictive than those
% in~\eqref{eq:qp1} since they are demanded to hold for all
% $i$. However, we will show that \eqref{eq:qp2} is feasible for a
% wide range of sets. We start the analysis of the all-components QP
% controller by a study of its regularity.
Our first result establishes regularity properties of the controller
$\uall$ if the program that defines it is feasible.

%
% \marginJC{Checking: so the argument here as in Th 4.1., but we don't
%   need the same hypotheses? It is true that Th 4.1 assumes stuff so
%   that we can guarantee the existence of the neighborhood, and here
%   instead we start from the existence as an hypothesis.}
%   \marginMA{Yes}
%
\begin{theorem}\label{thm:u2regular} {\rm \textbf{(Single Valued-ness
      and Continuity of All-Components QP Controller).}}
  Assume there exists a neighborhood $\Nc_{\partial \Cc}$ of
  $\partial \Cc$ where~\eqref{eq:qp2} is feasible for some $\alpha$
  and $\beta_i$'s.  Then, $\uall$ is single-valued. Furthermore, if
  $f$ and $G$ are continuous at $x \in \Nc_{\partial \Cc}$
  where~\eqref{eq:qp2} is strictly feasible,
  % (i.e., strict inequality applies),
  then $\uall$ is continuous at~$x$.
\end{theorem}
\begin{pf}
  Proving single valued-ness follows a similar argument as in the
  proof of Theorem~\ref{thm:discont}(ii).
  %
  % \marginJC{You mean to say that the same argument used to show
  % Theorem~\ref{thm:discont}(ii) can be used here? I think I'd say it
  % like that.}
  %
  As for continuity, keeping in mind the
  convexity of $\Uc$, the strict feasibility of~\eqref{eq:qp2}, and
  the continuity of $\nabla h_i$, $f$, $G$, $\alpha$, and $\beta$, the
  set-valued map defined by the constraints of~\eqref{eq:qp2} around $x$ is
  continuous by~\cite[Thms. 10 \& 12]{WWH:73}. Thus, $\uall$ is
  continuous at $x$ by~\cite[Cor. 8.1]{WWH:73}.  \oprocend
\end{pf}

Next, we show that $\uall$ is safe even at non-smooth points if the
program that defines it is feasible.
% \NA{``if feasible" is not precise here because if the program is not
% feasible, the controller is not defined.}.

\begin{theorem}\label{thm:u2safe} {\rm \textbf{(All-Components QP
      Controller Is Safe).}}
  Assume there exists a neighborhood $\Nc_{\partial \Cc}$ of
  $\partial \Cc$ where~\eqref{eq:qp2} is feasible for some $\alpha$
  and $\beta_i$'s. Then, any controller
  $k: \Cc \cup \Nc_{\partial \Cc} \to \Uc$ for which there exists a
  Filippov solution
  %
  % \marginJC{So far we've been writing consistently ``Filippov
  %   solution'', but not here. The inconsisteny might suggest to the
  %   reader that this is a different type of solution}
  %
  to the closed-loop dynamics from any point in $\Cc$ and
  that is equal to $\uall$ on $\Nc_{\partial \Cc}$ is safe with
  respect to~$\Cc$.
\end{theorem}
\begin{pf}
  By~\eqref{eq:Fil4Piece}, if $\zeta \in F[f+G\uall](x)$,
  $x \in \Nc_{\partial \Cc}$, then there exist $p$ sequences
  $\{x^1_\mu\}, \dots, \{x^p_\mu\}$ converging to $x$ such that
  \begin{align*}
    \zeta = \sum_{j = 1}^{p}\sigma_j \lim_{\mu \to \infty}(f(x^j_\mu)
    + G(x^j_\mu)\uall(x^j_\mu)), 
  \end{align*}
  where $\sigma_j$'s are convex combination constants. Thus, using the
  continuous differentiability of $h_i$, $i \in \until{r}$,
  \begin{align*}
    & \nabla h_i(x)^\top \zeta
    \\
    &= \sum_{j = 1}^{p}\sigma_j \lim_{\mu \to \infty} \nabla
      h_i(x^j_\mu)^\top(f(x^j_\mu) + G(x^j_\mu)\uall(x^j_\mu)). 
  \end{align*}
  Using~\eqref{eq:qp2} and taking the limit as $\mu \to \infty$, we
  obtain
  \begin{align*}
    & \nabla h_i(x_\mu^j)^\top (f(x^j_\mu) \!+\! G(x^j_\mu)\uall(x^j_\mu)) \geq
      - \alpha(h_i(x^j_\mu)) \!-\! \beta_i(H(x^j_\mu))
    \\ 
    & \implies \;\; \lim_{\mu \to \infty}\nabla h_i(x_\mu^j)^\top
      (f(x^j_\mu) + G(x^j_\mu)\uall(x^j_\mu))
    \\
    & \hskip 20mm \geq -\lim_{\mu \to \infty} (\alpha(h_i(x^j_\mu)) +
      \beta_i(H(x^j_\mu)))
    \\ 
    & \hskip 20mm = - \alpha(h_i(x))- \beta_i(H(x)).
  \end{align*}
  Therefore,
  $\nabla h_i(x)^\top \zeta \geq -\alpha(h_i(x))-\beta_i(H(x))$ for
  all $i \in \until{r}$. By~\eqref{eq:dh}, for any $v \in \partial h(x)$,
  $v = \sum_{i \in \Ic(x)}\bar \sigma_i\nabla h_i(x)$, where
  $\bar \sigma_i$'s are convex combination constants. Thus
  $v^\top \zeta \geq -\sum_{i \in \Ic(x)}\bar \sigma_i (\alpha(h_i(x))
  + \beta_i(H(x)))$. Recall that $\beta_i(H(x)) = 0$ when
  $i\in \Ic(x)$ and that if $i, i' \in \Ic(x)$ then
  $h(x) = h_i(x) = h_{i'}(x)$. Thus,
  $\sum_{i \in \Ic(x)}\bar \sigma_i (\alpha(h_i(x)) + \beta_i(H(x))) =
  \alpha(h(x))$, so we have $v^\top \zeta \geq -\alpha (h(x))$. The
  application of Theorem~\ref{thm:safetyCond} now completes the proof.
  \oprocend
\end{pf}

% So far, we established the regularity and safeness of $\uall$ so long
% as the program defining it is feasible.

We now turn our attention to establishing the feasibility of the
program defining~$\uall$. Specifically, we show that, under
Assumption~\ref{as:genCBFcond}, there exists a choice of $\alpha$ and
$\beta_i$'s that makes the program in \eqref{eq:qp2} feasible under
widely applicable conditions. Those conditions are summarized in the
following assumption.

\begin{assumption}\label{as:compactC} {\rm \textbf{(Structure of Safe
      Set).}}
  \rm{The set $\partial \Cc$ is compact and $\Cc$ is given by finitely
    many (arbitrary) nested unions and intersections of superlevel
    sets $\Cc_i \coloneqq \{x \in \real^n \;|\; h_i(x)\geq 0\}$ of the
    continuously differentiable functions $\{h_i\}_{i=1}^r$. We also
    assume that $h_i$'s are sufficiently different, i.e., for any
    $i \neq i'$, $h_i(x) \neq h_{i'}(x)$ for almost all $x \in \Xc$.
    % 
    % \marginJC{You mean there is a finite number of $h_i$'s? We already
    %   know, since we set $i \in \until{r}$. Or you mean that the same
    %   function can appear multiple times, but only a finite number of
    %   times?}
    } \oprocend
\end{assumption}
%
% \marginJC{If we say ``...finitely many (arbitrary) nested
%   unions...'', then I don't think we need the second sentence.}
%

The next result provides a normal form for any safe set that satisfies
Assumption~\ref{as:compactC}.
%
% \marginJC{This is a confusing sentence -- not clear what you want to
%   say. The definition IS mathematically precise. It is not fully
%   explicit. And this ``unlike what is done...'' refers to what? Why
%   the reference? What are we justifying? Why not simply say: The next
%   result provides a normal form for any safe set that satisfies
%   Assumption 2, or something like that.}
%

\begin{lemma}\label{lem:normalForm}{\rm \textbf{(Normal Form of Safe
      Set).}}
  Under Assumption~\ref{as:compactC}, $\Cc$ can be expressed as a
  union of intersections of collections of $\Cc_i$'s,
  \begin{align}
    \Cc = \bigcup_{\ell \in \Lc} \bigcap_{i \in \Ic^\ell} \Cc_i , \label{eq:CNormalForm}
  \end{align}
  with $\Lc $ finite and $\Ic^\ell \subseteq \{1,\dots,r\}$ for all
  $\ell \in \Lc$.
\end{lemma}
\begin{pf}
  Let $P_i$ denote the statement that $x \in \Cc_i$ and $P$ denote
  that $x \in \Cc$.  Let '$\lor$' and '$\land$' signify logical
  disjunction and conjunction, respectively. Then the statement that
  $x \in (\Cc_{i_1} \cup \Cc_{i_2})$ corresponds to
  $P_{i_1} \lor P_{i_2}$, while the statement that
  $x \in (\Cc_{i_1} \cap \Cc_{i_2})$ corresponds to
  $P_{i_1} \land P_{i_2}$.  Thus, by definition, $P$ is equivalent to
  a statement formed by nested applications of conjunction and
  disjunction on the elementary propositions~$P_i$.
  %
  % \marginJC{I think here you're missing saying something like
  %   ``union'' equals disjunction and ``intersection'' equals
  %   conjunction}
  %
  From~\cite[Part C, Central Thm.]{ELP:21}, the successive
  distribution of the conjunction over disjunction gives a statement
  that is equivalent to $P$ and is of the form of disjunction of
  conjunctions of collections of $P_i$'s. This means that $P$ is
  equivalent to
  \begin{align*}
    \bigvee_{\ell \in \Lc} \bigwedge_{i \in \Ic^\ell} P_i , 
  \end{align*}
  where $|\Lc|$ is the number of conjunction collections, and each
  $\Ic^\ell$ is a subset of $\{1,\dots,r\}$ containing all $i$ in the
  $j^{\text{th}}$ collection.
  % Thus, $x \in \Cc$ is equivalent to
  % $x \in \bigcup_{\ell \in \Lc} \bigcap_{i \in \Ic^\ell} \Cc_i$.
  \oprocend
\end{pf}

The previous result is not only an existence result but its proof also
provides guidance on how to explicitly represent a set $\Cc$ that
satisfies Assumption~\ref{as:compactC} as a union of intersections.
%
% \marginJC{Really? I don't see an algorithmic procedure to get
% it. Rather than this sentence, I'd say that the result provides
% guidance as to how to explicitly represent $\Cc$ in a way that is
% convenient for our purposes as a union of intersections.}
%
As a consequence of Lemma~\ref{lem:normalForm}, we have that the set
$\Cc$ corresponds to the $0$-superlevel set of the function
% Note that since any set $\Cc$ satisfying
% Assumption~\ref{as:compactC} can be expressed as a union of
% intersections
% $\bigcup_{\ell \in \Lc} \bigcap_{i \in \Ic^\ell} \Cc_i$,
\begin{align}
  h(x) = \max_{\ell \in \Lc} \min_{i \in \Ic^\ell} h_i(x) ,  \label{eq:hNormalForm}
\end{align}
where $h_i$ is the function that has $\Cc_i$ as its $0$-superlevel
set.  Next, we establish the feasibility of the program
defining~$\uall$.
% Now that a normal form is found for the sets $\Cc$, we are ready to
% prove the existence of functions $\alpha$ and $\beta_i$'s that will
% render the program defining $\uall$ in \eqref{eq:qp2} feasible for all
% safe sets satisfying Assumption~\ref{as:compactC}.

\begin{theorem}\longthmtitle{Feasibility of
    All-Components QP Controller}\label{thm:u2feasible}
  Under Assumption~\ref{as:compactC}, let
  $h^\ell(x) \coloneqq \min_{i \in \Ic^\ell}{h_i(x)}$ and define
  $ \Lc_i = \{\ell \in \Lc \;|\; i \in \Ic^\ell \}$. %\label{eq:Li}
  Let Assumption~\ref{as:genCBFcond} be satisfied with
  $\tilde \Ic(x) \coloneqq \{i \in \until{r} \;|\; \exists \ell \in
  \Lc_i,\; h(x) = h^\ell(x) = h_i(x)\}$.  Then, for any compact set
  $\Dc \subseteq \Cc$, there exist constants $\alpha$, $M \in \real$
  such that~\eqref{eq:qp2} is feasible in a neighborhood of $\Dc$ with
  %
  % \marginJC{We're using $r$ in $\until{r}$, and instead the $r$ in
  %   $\alpha$ is a completely different thing...}
  % 
  \begin{align}
    \alpha(\rho)
    & = \alpha \rho ,\notag
    \\
    \beta_i(H(x))
    & \coloneqq M \big
      (h(x)- \max_{\ell \in \Lc_{i}}h^\ell(x) \big
      ) .  \label{eq:beta}   
  \end{align}
  % 
  % \marginJC{Something is going on with this def: still does not make
  % sense: $\ell$ on the right is an arbitrary index, but in the
  % middle (for $\bar \beta_\ell$) is a specific one?}
  %
  % \NA{I think the statement of this theorem can be simplified. Why
  % define $\beta_\ell$ first and then $\beta_i$? Also, why repeat
  % twice that $\alpha(r) = \alpha r$? It is possible to directly
  % state that ``Let
  % $h^\ell(x) \coloneqq \min_{i \in \Ic^\ell}{h_i(x)}$. Under
  % Assumptions~\ref{as:genCBFcond} and~\ref{as:compactC}, there exist
  % constants $\alpha, M \in \real$ such that the program in
  % \eqref{eq:qp2} with $\alpha(r) = \alpha r$ and
  % $\beta_i(H(x)) = M(h(x)-h^\ell(x))$ for any $\ell$ with
  % $i \in \Ic^\ell$ is feasible." We should also clearly state that
  % $\mathcal{C}$ is of the form in Lemma \ref{lem:normalForm}.}
\end{theorem}
\begin{pf}
  We start by noting that, with the above definition,
  $ \Ic(x) \subseteq \tilde \Ic(x)$ for all $x \in \Xc$,
  cf. Lemma~\ref{lem:approxI}.  Let $\Dc \subseteq \Cc$ be an
  arbitrary compact set. For $i \in \tilde \Ic(x)$, we have
  $h(x) = \max_{\ell \in \Lc}h^\ell(x) = \max_{\ell \in
    \Lc_i}h^\ell(x)$.
  %
  % \marginJC{I don't see why. I only get
  % $h_i(x) \ge \max_{\ell \in \Lc_i}h^\ell(x)$} \marginMA{Good catch!
  % there was an unwarranted assumption that whenever
  % $h_i(x) = h^\ell(x)$ for some $i, \ell, x$ then $i \in \Ic^\ell$.}
  Thus $\beta_i(H(x)) = 0$ for all
  $i \in \Ic(x) \subseteq \tilde \Ic(x)$, ensuring the property of
  $\beta_i$ required in program~\eqref{eq:qp2}.  For convenience, we
  define $\bar \beta_\ell(H(x)) = M(h(x) - h^\ell(x))$ for
  $l \in \Lc$.  Next, we show that there exist $\alpha$ and $M$ such
  that, for all $x \in \Dc$, there is $u_x \in \Uc$ with
  $\nabla h_i(x)^\top (f(x) + G(x)u_x) + \alpha(h_i(x)) + \bar
  \beta_\ell (H(x)) > 0$, for all $\ell \in \Lc$ and all
  $i \in \Ic^\ell$.  This is sufficient for proving the statement
  since there exists $\ell \in \Lc_i$ such that
  $\beta_i (H(x))= \bar \beta_\ell (H(x))$.
  % \NA{Is the subscript on $\beta$ correct?}.
  Define first the indices set-valued map
  $L(x) \coloneqq \{\ell \in \Lc \;|\; h(x) = h^\ell(x)\}$. Our proof
  has three steps.

  \textbf{Step 1:} Here, we prove there exists $\alpha \in \real_{>0}$
  such that
  \begin{align*}
    &\forall x \in \Dc, \; \exists u_x 
      \in \Uc \text{ with } 
    \\ 
    & \hskip 10mm \nabla h_i(x)^\top(f_j(x)+G_j(x)u_x) + \alpha h_i(x) > 0,
  \end{align*}
  for all $\ell \in L(x)$, $ i \in \Ic^\ell$, and $ j \in \Jc(x)$.
  This is sufficient for the satisfaction of all constraints on $h_i$
  with $i \in \Ic^\ell$ and $\ell \in L(x)$.  It is even stronger than
  mere feasibility of the constraint of $h_i$ at $x$ since we ask for
  the constraint to be satisfied for all $j \in \Jc(x)$ (we use this
  in Step~3 of the proof). We reason by contradiction.  Suppose that
  this is not the case. Then, for all $n \in \mathbb{N}$, there exists
  $x_n \in \Dc$ such that for all $ u_x \in \Uc$, there are
  $ \ell \in L(x_n)$, $ i \in \Ic^\ell$, and $ j \in \Jc(x_n)$ with
  \begin{align}
    \nabla h_i(x_n)^\top (f_j(x_n)+G_j(x_n)u_x) + n h_i(x_n)
    \leq 0. \label{eq:4contradiction}
  \end{align}
  Without loss of generality, due to the compactness of $\Dc$, the
  sequence $\{x_n\}$ is convergent~\cite[Thm. 3.6]{WR:76}, say to
  $\bar x \in \Dc$.  Let
  $I_{\bar x} = \setdef{i \in \until{r}}{\exists \ell \in L(\bar x)
    \text{ s.t. } i \in \Ic^\ell \text{ and } h_i(\bar x) = 0 }$. Note
  that $I_{\bar x} \subseteq {\tilde \Ic(\bar x)}$, since, for
  $\ell \in L(\bar x)$,
  $h(\bar x) = h^\ell(\bar x) = \min_{i\in \Ic^\ell}h_i(x) \geq 0$,
  which gives $h(\bar x) = h^\ell(\bar x)= h_i(\bar x) = 0$ for
  $i \in I_{\bar x}$.
  %
  % \marginJC{Are you trying to say that
  % $I_{\bar x} \subset \Ic(\bar x)$?}  \marginMA{this will do too.}
  %
  Thus, by Assumption~\ref{as:genCBFcond}, there is $u_{\bar x}$ that
  validates
  \begin{align}\label{eq:ubarx}
    \nabla h_{i}(\bar x)^\top(f_j(\bar x)+G_j(\bar x)u_{\bar x}) > 0,
  \end{align}
  % 
  % \marginJC{This is true b/c $h_{i}(\bar x)=0$, correct?}
  % \marginMA{and because $i \in \Ic^\ell$ and $\ell \in L(x)$. In
  % short, because $I_{\bar x} \subset \Ic^\ell$ as you suggested.}
  for all $i \in I_{\bar x}$ and $j \in \Jc(\bar x)$. If $I_{\bar x}$
  is empty, take $u_{\bar x}$ as any finite value in
  $\Uc$. By~\eqref{eq:4contradiction}, with the choice
  $u_x = u_{\bar x}$, for every $n$ there is
  $\ell_n \in L(x_n) \subseteq \Lc$, $i_n \in \Ic^{\ell_n}$ and
  $j_n \in \Jc(x_n)$ that validate~\eqref{eq:4contradiction}.
  Since $\ell_n, i_n,$ and $j_n$ are in finite sets they converge
  without loss of generality~\cite[Thm. 3.6]{WR:76}, say to
  $\bar \ell, \bar i,$ and $\bar j$ respectively. Hence, for $n$ large
  enough, we have $\bar \ell \in L(x_n)$,
  $\bar i \in \Ic^{\bar \ell}$, $\bar j \in \Jc(x_n)$ and
  \begin{align}
    \nabla h_{\bar i}(x_n)^\top (f_{\bar j}(x_n)+G_{\bar
    j}(x_n)u_{\bar x}) + nh_{\bar i}(x_n) 
    \leq 0. \label{eq:4contradiction2}
  \end{align} 
  Because $\bar \ell \in L(x_n)$,
  $h^{\bar \ell}(x_n) = \min_{i \in \Ic^{\bar \ell}} h_i(x_n) = h(x_n)
  \geq 0$. Therefore, we deduce
  $h_{\bar i}(x_n) \geq h^{\bar \ell}(x_n) \geq 0$. The continuity of
  $\nabla h_{\bar i}(\cdot)^\top (f_{\bar j}(\cdot)+G_{\bar
    j}(\cdot)u_{\bar x})$ in a neighborhood of $\bar x$ implies that,
  for sufficiently large $n$, this expression evaluated at $x_n$ is
  bounded.  This, together with~\eqref{eq:4contradiction2} and the
  fact that $h_{\bar i}(x_n) \geq 0$ implies that
  $h_{\bar i}(\bar x)=0$.  By Lemma~\ref{lem:ij},
  $\bar \ell \in L(x_n) \subseteq L(\bar x)$ and
  $\bar j \in \Jc(x_n) \subseteq \Jc(\bar x)$ and thus
  $\bar i \in I_{\bar x}$. Thus~\eqref{eq:ubarx} with $i = \bar i$ and
  $j = \bar j$ contradicts~\eqref{eq:4contradiction2}.

  \textbf{Step 2:} Let $\alpha$ satisfy the statement of Step~1. Here,
  we prove that there exists $M > 0$ such that, for all $x \in \Dc$,
  there is $u_x$ validating
  $\nabla h_i(x)^\top (f_j(x) + G_j(x)u_x) + \alpha(h_i(x)) + \bar
  \beta_\ell(H(x)) > 0$ for all $\ell \in \Lc$, $i \in \Ic^\ell$, and
  $j \in \Jc(x)$.  We reason again by contradiction and assume this is
  not the case. Then, for all $n$, there exists $x_n \in \Dc$ such that,
  for all $u \in \Uc$, there are $\ell \in \Lc$, $i \in \Ic^\ell$, and
  $ j \in \Jc(x_n)$ with
  \begin{multline}\label{eq:4contradictionM}
    \nabla h_i(x_n)^\top (f_j(x_n) + G_j(x_n)u) +
    \alpha(h_i(x_n)) 
    \\
    + n(h(x_n) - h^\ell(x_n)) \leq 0 . 
  \end{multline}
  Again, without loss of generality, the sequence $x_n$
  converges~\cite[Thm. 3.6]{WR:76}, say to $\bar x \in \Dc$. By Step 1
  and the fact that $h(x) \geq h^\ell(x)$ for all $\bar x \in \Dc$, there
  exists $u_{\bar{x}}$
  %
  % \marginJC{Shouldn't this be $u_{\bar{x}}$?}
  % %
  % \marginMA{No, it is the $u_x$ that appears in the first inequality
  %   in Step 1, so it is consistent. (Anyway it is a bound variable so
  %   it does not matter as you know.)}
  % %
  % \marginJC{Right, but we are invoking Step 1 for the specific state
  %   $\bar{x}$, hence $u_{\bar x}$. $x$ is arbitrary, so $u_x$ does not
  %   make sense.}
  %
  such that, for all $\ell \in L(\bar x)$,
  $i \in \Ic^\ell$, and $j \in \Jc(\bar x)$,
  \begin{align*}
    \nabla h_i(\bar x)^\top (f_j(\bar x) \!+\! G_j(\bar x)u_{\bar{x}}) \!+\!
    \alpha(h_i(\bar x)) \!+\! n(h(\bar x) \!-\! h^\ell(\bar x)) > 0 .    
  \end{align*}
  Due to continuity, Lemma~\ref{lem:ij}, and the fact that
  $h(x) \geq h^\ell(x)$ for all $x \in \Dc$ and $\ell \in \Lc$, there
  is a neighborhood $B_\epsilon(\bar x)$ such that
  \begin{align*}
    \nabla h_i(y)^\top (f_j(y) \!+\! G_j(y)u_{\bar{x}}) \!+\!
    \alpha(h_i(y)) \!+\! n(h(y)
    \!-\! h^\ell(y)) > 0 , 
  \end{align*}
  for all $n > 0$, $y \in B_\epsilon(\bar x)$,
  % % 
  % \marginJC{You mean $y \in B_\epsilon(\bar x)$?}
  % % 
  $\ell \in L(\bar x)$, $i \in \Ic^\ell$, and $j \in \Jc(y)$. Thus,
  for large enough $n$, there exists $u_{\bar{x}}$ that falsifies
  \eqref{eq:4contradictionM} for all $\ell \in L(\bar x)$,
  $i \in \Ic^\ell$, and $j \in \Jc(x_n)$.  For
  $\ell \notin L(\bar x)$, $n(h(x_n) - h^\ell(x_n)) \to \infty$. But
  continuity ensures boundedness of
  $\nabla h_i(y)^\top (f_j(y) + G_j(y)u_{\bar{x}}) + \alpha(h_i(y))$ for
  $y \in B_\epsilon(\bar x)$. Therefore, for large enough
  $n$,~\eqref{eq:4contradictionM} is false for all
  $\ell \notin L(\bar x)$, $i \in \Ic^\ell$, and $j \in
  \Jc(x_n)$. Therefore, we have shown that~\eqref{eq:4contradictionM}
  is falsified for large enough $n$ for all $\ell \in \Lc$,
  $i \in \Ic^\ell$, and $j \in \Jc(x_n)$, completing the proof of
  Step~2.

  \textbf{Step 3:}
  % Step 2 proves the existence of $\alpha$ and $M$ that validate the
  % strict inequalities of the constraints of the
  % program~\eqref{eq:qp2} in the compact set $\Dc$ for all
  % $j \in \Jc(x)$.
  In this step, we prove that constraint satisfaction established in
  Step 2 with $\alpha$ and $M$ is valid on a neighborhood of
  $\Dc$. For all points $x \in \partial \Dc$, Step~2 establishes the
  existence of $u_x$ that validates the strict inequality constraints
  of the program~\eqref{eq:qp2} defining $\uall$ for all
  % 
  % \marginJC{Argh! $\ell \in \Lc$ still missing here?}
  % %
  % \marginMA{I meant the constraints of $\uall$ which are independent
  %   of $\ell$. My bad for saying $i \in \Ic^\ell$.}
  % %
  % \marginJC{But this relies on something like ``$\ell \in \Lc$ and
  %   $i \in \Ic_\ell$'' is the same as $i \in \until{r}$. We don't
  %   quite make that explicit in Lemma 4.8.}
  % 
  $i \in \cup_{\ell \in \Lc}\Ic^\ell$ and $j \in \Jc(x)$. By
  continuity and Lemma~\ref{lem:ij}, there exists a neighborhood of
  $x$ in which $u_x$ validates the constraints. \oprocend
\end{pf}
% We mentioned previously that the functions $\beta_i$ in the definition of $\uall$ will be designed to to facilitate the satisfaction of the constraints, i.e., feasibility. The design of $\beta_i$ given in~\eqref{eq:beta} does this by making $\beta_i(H(x)) > 0$ whenever $h_i(x) < 0$ and parameterizing it by $M$. So whenever $h_i(x) > 0 $ 
% \marginJC{So far, we've never got back in the exposition to the
%   earlier comment that 'beta's will facilitate satisfaction of
%   contraints''. We should comment on it here, saying that the design
%   in (9) makes it easier b/c $\beta$ takes positive values and bla,
%   bla.}

\begin{figure}
  \centering
  \begin{tikzpicture}
    \node at (0,0) {\includegraphics[width=.3 \textwidth]{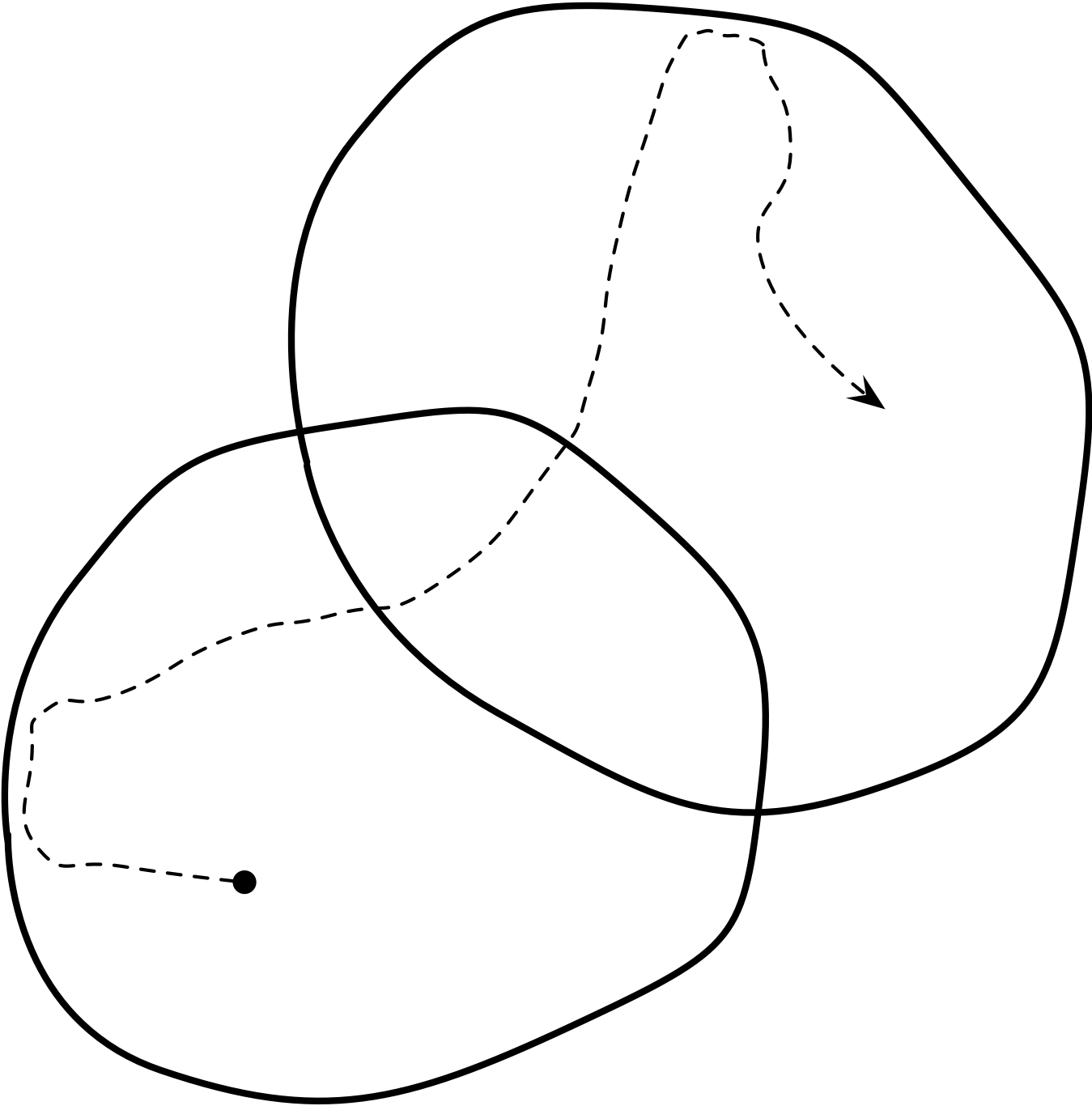}};
    \draw (-0.8,-1.5) node {$\Cc_1$};
    \draw (-0.8,-2) node {$\beta_1(x) = 0$};
    \draw (1,0.3) node {$\Cc_2$};
    \draw (0.9,-0.2) node {$\beta_1(x) > 0$};
  \end{tikzpicture}
  \caption{Illustration of the role of the transition function. The
    safe set is $\Cc_1 \cup \Cc_2$. The continuous function $\beta_1$ is positive
    at the boundary points of $\Cc_1$ only if they are in $\interior{\Cc_2}$, and zero otherwise in $\Cc_1$. The
    constraints in the all-components QP controller $\uall$ defined
    in~\eqref{eq:qp2} allow trajectories to leave one set only to the
    other, while remaining in the safe set.}
  \label{fig:beta}
\end{figure}
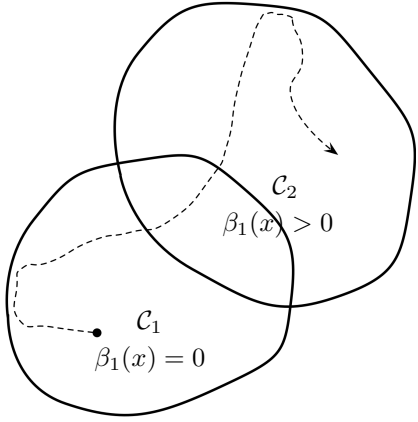

\begin{remark}\longthmtitle{Role of Transition
    Functions}\label{rem:constraintsMeaning} {\rm To understand the
    role of the transition function $\beta_1$, consider the simple case where
    the safe set is given by a superlevel set $\Cc_1$ of one
    differentiable function $h_1$.  The standard CBF
    constraint~\eqref{eq:cbfCond} for $h_1$ ensures that
    $h_1(x(t)) \geq 0$, and hence the forward invariance
    of~$\Cc_1$. This constraint is altered in our proposed
    design~\eqref{eq:qp2} as
    \begin{equation*}
      \nabla h_1(x)^\top(f(x)+G(x)u)
      + \alpha(h_1(x)) + \beta_1(x) \geq 0 .
    \end{equation*}
    Note that,
    under this constraint, system trajectories will not leave $\Cc_1$
    from points $x\in \partial \Cc_1$ where $\beta_1(x)\le 0$,
    cf. Theorem~\ref{thm:safetyCond}, whereas such a guarantee does not
    exist from points $x\in \partial \Cc_1$ where $\beta_1(x) > 0$. A
    nonnegative transition function therefore makes it easier to
    satisfy the constraint while possibly inducing the violation of
    the forward invariance of $\Cc_1$ from boundary points where the
    function is strictly positive.  This is of course not desired if
    the goal is to keep $\Cc_1$ safe. However, if our safety
    requirement is keeping a union $\Cc_1 \cup \Cc_2$ safe and
    $\beta_1$ is designed to allow the trajectories to leave $\Cc_1$
    from points of the boundary which are in $\Cc_2$,
    cf. Figure~\ref{fig:beta}, then the constraint with $\beta_1$
    provides the flexibility to ``transition'' from $\Cc_1$ to
    $\Cc_2$. This idea is extended in our design~\eqref{eq:qp2} to
    deal with multiple component sets, where the transition functions
    are chosen to be positive where it is desirable to allow the
    trajectory to leave one of those component sets to get to
    another. The idea of relaxing safety for some parts of a set might
    be of interest in other contexts, such as prescribed time and
    space safety, where trajectories are to be kept in a set for some
    time and then allowed to exit only from specific points. In such
    cases, the transition functions might depend on time as well as on
    space.
    % An
    % exploration of the possibilities that the inclusion of $\beta$
    % open are left for future works. Now that the use of $\beta$ is
    % understood, note that the greater the constant $M$ in our choice
    % of $\beta$ in Theorem~\ref{thm:u2feasible} in, the greater $\beta$
    % is for any $x$. Thus, choosing $M$ large enough enhances the
    % possibility of trajectories leaving from where we want them to
    % leave. Moreover, one should be cautious when designing the
    % component sets whose union comprises $\Cc$. If the components
    % intersect in sets with empty interiors then no matter how large
    % $M$ is in our choice of $\beta$, the trajectories will not be able
    % to leave the component sets since $\beta(x) = 0$ for all boundary
    % points of such a component. This will not violate safety but will
    % hinder the satisfaction of any additional control objectives that
    % require leaving one of the component sets to another.  
  } \oprocend
\end{remark}

%
% \marginJC{Shouldn't this be $\beta_1$ and $\beta_2$ instead of just
%   $\beta$? Also, in the plot, could you draw trajectories in say
%   $\Cc_1$ that go towards its boundary (but not towards $\Cc_2$) and
%   then get pulled back into $\Cc_1$, and trajectories that start in
%   $\Cc_1$ and cross over to $\Cc_2$?}
%\marginMA{I don't know why the arrow tip does not show up at the end of the trajectory. Same with you?}
%
%
% \marginJC{Right, I don't see the arrow tip. Also, I'd like to see
%   another trajectory that stays in $\Cc_1$}
% %

\begin{remark}\longthmtitle{Properties of the Feasibility Set}
  \label{rem:FeasibilityExtended}
  \rm{From the proof of Theorem~\ref{thm:u2feasible}, one can see that
    if $\alpha$ satisfies the statement of Step 1, then any
    $\bar \alpha > \alpha$ does too. Similarly, if $M$ satisfies Step
    2 for some $\alpha$, then for any $\bar \alpha > \alpha$, there
    exists $\bar M > M$ satisfying Step 2. Thus, for any pair
    $(\alpha, M)$ satisfying Theorem~\ref{thm:u2feasible}, there is a
    pair of greater numbers $(\bar \alpha, \bar M)$ that do too. We
    leverage this observation below in our approach to find
    appropriate parameters $\alpha$ and $M$ that ensure feasibility.
    A larger value of $\alpha$ allows for faster approaches to the
    boundary of the safe set~\cite{ADA-SC-ME-GN-KS-PT:19}, while a
    larger $M$ enhances the ability to transition from one component
    of the safe set to another
    (cf. Remark~\ref{rem:constraintsMeaning}).  } \oprocend
\end{remark}
%
% \marginJC{What does a controller with a larger $\alpha$ mean? How
%   about a controller with a larger $M$?}
% %

\subsection{All-Components Adaptive QP Controller}

Our design in this section is motivated by the observation that
Theorem~\ref{thm:u2feasible} is an existence result, but not a
constructive one, in the sense that it does not provide an explicit
way to construct the functions $\alpha$ and $\beta_i$'s that ensure
the feasibility of the program defining~$\uall$.  We remedy this difficulty by
introducing a controller where $\alpha$ and $M$ (which in turn defines
the functions $\beta_i$'s) are taken as optimization variables
themselves. Namely, using the fact that
$h (\cdot) = \max_{\ell \in \Lc} \min_{i \in \Ic^\ell} h_i(\cdot)$
(cf. Lemma~\ref{lem:normalForm}), we propose the following design:
\begin{align}
  (\uadp(x),
  &\alphaadp(x),\Madp(x)) \coloneqq \notag
  \\
  & \argmin_{\alpha, M, u \in \Uc} {u^\top Q(x)u + b(x)^\top u +
    q_\alpha \alpha^2 + q_M M^2} \nonumber
  \\
  \text{s.t. } & M \geq c_M, \; \alpha \geq c_\alpha \label{eq:qp3}
  \\
  &\nabla h_i(x)^\top (f(x) + G(x)u) + \alpha h_i(x) \nonumber
  \\
  &+ M(h(x) - h^\ell(x)) \geq 0,  \; \forall \ell \in \Lc, \;
    i \in \Ic^\ell, \nonumber 
\end{align}
where $c_M, c_\alpha, q_\alpha, q_M$ are positive constants
% \marginJC{What's the role of these constants? We can say that the
% $q$'s make sure to penalize large $\alpha$, $M$'s (and connect with
% whatever we add in response to my margin above). What's the
% meaning/impact of $c_M, c_\alpha$?}
%
and the rest of the conditions on the objective function are the same
as in~\eqref{eq:qp1}.  The positive constants $q_\alpha$ and $q_M$
establish the positive definiteness of the objective function, which
ensures the uniqueness of the solver of~\eqref{eq:qp3}. They also
penalize large $\alpha$ and $M$, and thus establish the existence of
an upper bound for $\alphaadp$ and $\Madp$ in any compact set. This is
needed for establishing control invariance. The positivity of
$c_\alpha$ keeps the quantity $\alphaadp(x)h_i(x)$ greater than the
class-$\kappa$ function $\alpha(h(x)) = c_\alpha h_i(x)$, a fact
necessary for establishing control invariance. The positivity of $c_M$
ensures that $\Madp(x)(h(x) - h^\ell(x))$ is positive when
$h(x) \neq h^\ell(x)$. In this way, it promotes violating the safety
constraint of the inactive components of $h$, which is necessary for
transitioning between different components of the safe set.  Since the
values of $\alpha$ and $M$ are chosen adaptively in~\eqref{eq:qp3} as
a function of the state, we refer to this as the \emph{all-components
  adaptive QP controller}.
% \NA{This name has a different order than the name of the
% subsection. Make these consistent.}.

\begin{theorem}{\rm \textbf{(Properties of All-Components Adaptive QP
      Controller).}}\label{thm:adp}
  Under the assumptions of Theorem~\ref{thm:u2feasible}, in any
  compact $\Dc \subseteq \Cc$, the adaptive all-components QP
  controller~$\uadp$ defined by~\eqref{eq:qp3} is
  \begin{enumerate}[(i)]
    \item feasible,
    \item single-valued,
    \item continuous wherever the
    system dynamics is continuous, and
      %
      % \marginJC{Wouldn' it be better to say ``continuous wherever the
      %   system dynamics is continuous''?}
      %
    \item safe, i.e., no trajectory can leave $\Cc$ from a point in $\Dc$.
  \end{enumerate}
\end{theorem}
\begin{pf}
  Statement~(i) follows from Theorem~\ref{thm:u2feasible} and
  Remark~\ref{rem:FeasibilityExtended}.  Statement~(ii) can be
  established with the same argument as in
  Theorem~\ref{thm:discont}(ii) and statement (iii) with the same
  argument as in Theorem~\ref{thm:u2regular}.  Regarding statement
  (iv), let $\bar M$ and $\bar \alpha$ be upper bounds of $\Madp(x)$
  and $\alphaadp(x)$ on $\Dc$.
  %
  % \marginJC{Imprecise: you mean upper bounds on $\Dc$? Also, ``surely
  %   exist'' might be confusing (why not simply ``exist''?): somebody
  %   could think you're talking about ``a.e.''}
  %
  Those upper bounds exist since by
  Theorem~\ref{thm:u2feasible}, finite $\alpha$ and $M$ exist that
  validate the constraints for all $x \in \Dc$ at once. Now define the
  class-$\kappa$ function $\tilde \alpha(r) = \bar \alpha r$ for
  $r \geq 0$ and $\tilde \alpha(r) = c_\alpha r$ for $r < 0$. Then,
  $\uadp$ satisfies the constraints of~\eqref{eq:qp2} with
  $\alpha(r) = \tilde \alpha(r)$ and
  $\beta_i(H(x)) = \bar M \big (h(x)- \max_{\ell \in \Lc_{i}}h^\ell(x)
  \big )$, with $\Lc_i$ defined in Theorem~\ref{thm:u2feasible}, because, for each
  $x \in \Dc$,
  \begin{multline*}
    \tilde \alpha(h(x)) + \bar M \big (h(x) - \max_{\ell \in
      \Lc_{i}}h^\ell(x) \big ) \geq
    \\
    \alphaadp(x)h(x) + \Madp(x)\big
    (h(x) - \max_{\ell \in \Lc_{i}}h^\ell(x) \big ).
  \end{multline*}
  %
  % \marginJC{just before equation: ``for each $x \in \Dc$''; just after
  %   ``in $\Dc$".}
  %
  With the argument as in Step 3 in the proof of
  Theorem~\ref{thm:u2feasible}, we deduce that the constraints are
  satisfied
  %
  % \marginJC{``are valid''... You mean, ``are satisfied''?}
  % 
  with $\tilde \alpha$ and $\bar M$ on a neighborhood of $\Dc$. Hence,
  the same reasoning employed to show Theorem~\ref{thm:u2safe} ensures
  $\uadp$ is safe. \oprocend
\end{pf}

The all-components QP controller $\uall$ and its adaptive version
$\uadp$ provide solutions to Problem~\ref{prob:getController}.  Given
discontinuous dynamics and safety constraints defined by multiple
components, we provide controllers that are safe, computable online,
and continuous wherever the dynamics is continuous, under standard
safety assumptions. We illustrate the versatility of the proposed
designs in a multi-agent reconfiguration problem in
Section~\ref{sec:simulation} below.

\section{Application to Multi-Agent Reconfiguration Control
  Problem}\label{sec:simulation}

\begin{figure*}[ht!]
  \centering
  \begin{subfigure}{0.33 \linewidth} 
    \centering
    \resizebox{\linewidth}{!}{
      \begin{tikzpicture}
      \begin{axis}[]
        \draw[draw = gray, fill = gray] (axis cs: 0,.1) rectangle (axis cs: -5,10);
        \addplot[black, only marks, mark = o, mark size = 7pt, mark options={fill = black}] table [x=xi,y=yi, col sep=comma]{multiagentEx_Data.csv};        
        \addplot[dotted, only marks, black, mark = o, mark size=7pt] table [x=x11s1, y=x12s1, col sep=comma] {multiagentEx_Data.csv};
        \addplot[dotted, black, very thick] table [x=x11s1, y=x12s1, col sep=comma] {multiagentEx_Data.csv};
        \addplot[only marks, mark = o,black, dotted,mark size=7pt] table [x=x21s1, y=x22s1, col sep=comma] {multiagentEx_Data.csv};
        \addplot[dotted, black, very thick] table [x=x21s1, y=x22s1, col sep=comma] {multiagentEx_Data.csv};
        \addplot[only marks, mark = o,black, dotted,mark size=7pt] table [x=x31s1, y=x32s1, col sep=comma] {multiagentEx_Data.csv};
        \addplot[dotted, black, very thick] table [x=x31s1, y=x32s1, col sep=comma] {multiagentEx_Data.csv};
        \addplot[only marks, mark = o,black, dotted,mark size=7pt] table [x=x41s1, y=x42s1, col sep=comma] {multiagentEx_Data.csv};
        \addplot[dotted, black, very thick] table [x=x41s1, y=x42s1, col sep=comma] {multiagentEx_Data.csv};
        \addplot[only marks, mark = o,black, dotted,mark size=7pt] table [x=x51s1, y=x52s1, col sep=comma] {multiagentEx_Data.csv};
        \addplot[dotted, black, very thick] table [x=x51s1, y=x52s1, col sep=comma] {multiagentEx_Data.csv};
        \addplot[black, only marks, mark = *, mark size = 7pt, mark options={fill = black}] table [x=xfs1,y=yfs1, col sep=comma]{multiagentEx_Data.csv};
        \addplot[black, only marks, mark = x] table [x=xf,y=yf, col sep=comma]{multiagentEx_Data.csv};
      \end{axis}
    \end{tikzpicture}
  }
\end{subfigure}
\hfill
\begin{subfigure}{0.33 \linewidth} 
  \centering
  \resizebox{\linewidth}{!}{
    \begin{tikzpicture}
      \begin{axis}[]
        \draw[draw = gray, fill = gray] (axis cs: 0,0.1) rectangle (axis cs: -5,10);
        \addplot[black, only marks, mark = o, mark size = 7pt, mark options={fill = black}] table [x=xi,y=yi, col sep=comma]{multiagentEx_Data.csv};        
        \addplot[dotted, only marks, black, mark = o, mark size=7pt] table [x=x11s2, y=x12s2, col sep=comma] {multiagentEx_Data.csv};
        \addplot[dotted, black, very thick] table [x=x11s2, y=x12s2, col sep=comma] {multiagentEx_Data.csv};
        \addplot[only marks, mark = o,black, dotted,mark size=7pt] table [x=x21s2, y=x22s2, col sep=comma] {multiagentEx_Data.csv};
        \addplot[dotted, black, very thick] table [x=x21s2, y=x22s2, col sep=comma] {multiagentEx_Data.csv};
        \addplot[only marks, mark = o,black, dotted,mark size=7pt] table [x=x31s2, y=x32s2, col sep=comma] {multiagentEx_Data.csv};
        \addplot[dotted, black, very thick] table [x=x31s2, y=x32s2, col sep=comma] {multiagentEx_Data.csv};
        \addplot[only marks, mark = o,black, dotted,mark size=7pt] table [x=x41s2, y=x42s2, col sep=comma] {multiagentEx_Data.csv};
        \addplot[dotted, black, very thick] table [x=x41s2, y=x42s2, col sep=comma] {multiagentEx_Data.csv};
        \addplot[only marks, mark = o,black, dotted,mark size=7pt] table [x=x51s2, y=x52s2, col sep=comma] {multiagentEx_Data.csv};
        \addplot[dotted, black, very thick] table [x=x51s2, y=x52s2, col sep=comma] {multiagentEx_Data.csv};
        \addplot[black, only marks, mark = *, mark size = 7pt, mark options={fill = black}] table [x=xfs2,y=yfs2, col sep=comma]{multiagentEx_Data.csv};
        \addplot[black, only marks, mark = x] table [x=xf,y=yf, col sep=comma]{multiagentEx_Data.csv};
      \end{axis}
    \end{tikzpicture}  
  }
  \end{subfigure}
  \hfill
  \begin{subfigure}{0.33 \linewidth} 
    \centering
    \resizebox{\linewidth}{!}{
      \begin{tikzpicture}
      \begin{axis}[]
        \draw[draw = gray, fill = gray] (axis cs: 0,.1) rectangle (axis cs: -5,10);
        \addplot[black, only marks, mark = o, mark size = 7pt, mark options={fill = black}] table [x=xi,y=yi, col sep=comma]{multiagentEx_Data.csv};        
        \addplot[dotted, only marks, black, mark = o, mark size=7pt] table [x=x11, y=x12, col sep=comma] {multiagentEx_Data.csv};
        \addplot[dotted, black, very thick] table [x=x11, y=x12, col sep=comma] {multiagentEx_Data.csv};
        \addplot[only marks, mark = o,black, dotted,mark size=7pt] table [x=x21, y=x22, col sep=comma] {multiagentEx_Data.csv};
        \addplot[dotted, black, very thick] table [x=x21, y=x22, col sep=comma] {multiagentEx_Data.csv};
        \addplot[only marks, mark = o,black, dotted,mark size=7pt] table [x=x31, y=x32, col sep=comma] {multiagentEx_Data.csv};
        \addplot[dotted, black, very thick] table [x=x31, y=x32, col sep=comma] {multiagentEx_Data.csv};
        \addplot[only marks, mark = o,black, dotted,mark size=7pt] table [x=x41, y=x42, col sep=comma] {multiagentEx_Data.csv};
        \addplot[dotted, black, very thick] table [x=x41, y=x42, col sep=comma] {multiagentEx_Data.csv};
        \addplot[only marks, mark = o,black, dotted,mark size=7pt] table [x=x51, y=x52, col sep=comma] {multiagentEx_Data.csv};
        \addplot[dotted, black, very thick] table [x=x51, y=x52, col sep=comma] {multiagentEx_Data.csv};
        \addplot[black, only marks, mark = *, mark size = 7pt, mark options={fill = black}] table [x=xf,y=yf, col sep=comma]{multiagentEx_Data.csv};
      \end{axis}
    \end{tikzpicture}
    }
  \end{subfigure}
  \caption{Illustration of the adaptive all-components controller
    $\uadp$ acting in a multi-agent reconfiguration problem. From left
    to right, different snapshots of the agent evolution as time
    progresses are portrayed.  The agents (black circles) travel from initial points
    (empty circles) to final destinations (crosses) while avoiding the
    unsafe region (gray area) and colliding with each other.  This
    illustrates how the proposed control design handles nested
    disjunctive and conjunctive constraints.}
  \label{fig:multiagentEx}
\end{figure*}
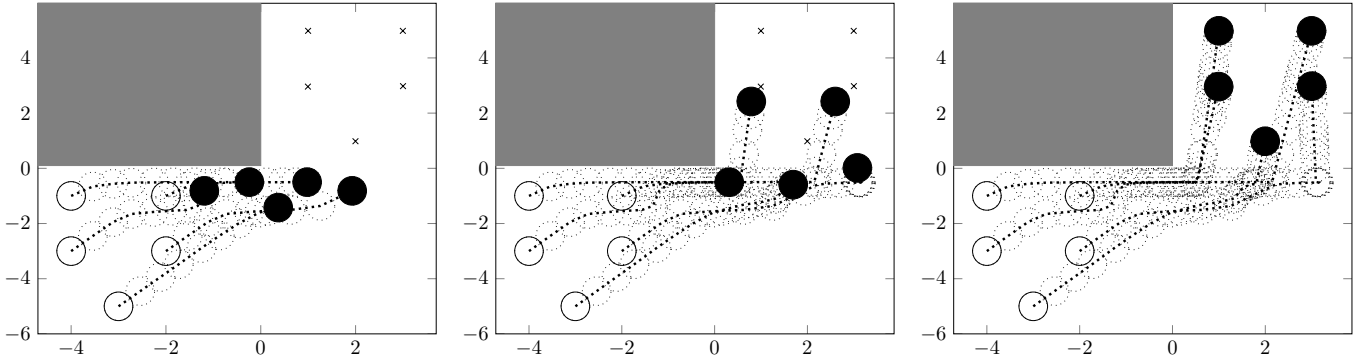
%
% \marginJC{Would it make sense to also plot the evolution of
%   $\beta_i$'s as a function of time? That could give us a chance to
%   speak about when the agent evolution ``transitioned'' from one safe
%   set to another.}
% \marginMA{This is generally a good idea. But for this example it is quite messy. The reason is that $\Lc = \{1,\dots, 32\}$. So we have $32$ components whose union is the safe set. Also those components are in the state-space which is here $(\real^2)^5 = \real^10$. Thus, we have $32$ $\bar beta_\ell$'s, each for a component. The components might not be easily visualizable sets in our $2D$ figure. I can introduce ad hoc sets of two $\bar \beta_\ell$'s, one roughly corresponds to the area $x_{i,2} < -\delta$ and another to $x_{i,1} > \delta$ for some agent $i$ and show that for some time in the begining the first is $0$ while the other positive, and the opposite happens as time passes indicating transition. But it won't  be precise.}
%

In this section, we demonstrate the efficacy of our control design in
a multi-agent scenario inspired by~\cite{PG-JC-ME:17-csl}. We consider
$5$ robot agents with positions $x_i = (x_{i,1},x_{i,2})$,
$i \in \until{5}$, and first-order dynamics $\dot x_i = u_i$.  Define
the full state as $x = (x_1, \dots, x_5)$. We require our agents to
travel from start points to end points with the safety requirements of
staying in the obstacle-free space and not colliding with each
other. Taking $\delta$ to be the agents' radius, we define the
obstacle-free space by the requirement that
\begin{align*}
  \text{either} \;
  x_{i,1} - \delta \geq 0 \quad \text{or} \quad  x_{i,2} + \delta \leq 0,
\end{align*}
for all $i \in \until{5}$. Agents $i$ and $j$ are not in collision so
long as
$\bar h_{i,j}(x_i,x_j) = \|x_i - x_j\|^2 - (2\delta)^2 \geq
0$. Denoting
$\Cc_{i,1} = \{x = (x_1,x_2)\;|\;h_{i,1}(x) = x_{i,1}-\delta \geq 0\}$,
$\Cc_{i,2} = \{x \;|\;h_{i,2}(x) = -x_{i,2} - \delta \geq 0\}$,
%
% \marginJC{The 1 in $x_{i,1}-1$ and $-x_{i,2} - 1$ should be $\delta$, no?}
%
and $\bar \Cc_{i,j} = \{x\;|\; \bar h_{i,j}(x) \geq 0\}$, the safe set
is given by
$\Cc = (\cap_{i}(\Cc_{i,1} \cup \Cc_{i,2})) \cap (\cap_{i\neq j}\bar
\Cc_{i,j})$. By the algebraic manipulation employed in
Lemma~\ref{lem:normalForm}, $\Cc$ can be put in the form of a union of
intersections $\cap_{\ell \in \Lc} \cup_{i\in \Ic^\ell} \Ac_i$, where
each $\Ac$ is either $\Cc_{i,1}, \Cc_{i,2}$ or $\bar \Cc_{i,j}$. We
omit the details of this transformation in the interest of
brevity. This representation allows us to apply the adaptive
all-components QP controller~\eqref{eq:qp3}. Since we want the agents
to travel from start point $x_s$ to final point $x_f$, we use
$\|u - u_{\nom}(x)\|^2$ for the objective function in~\eqref{eq:qp3},
with nominal controller $u_{\nom}(x) = x_f -
x$. Figure~\ref{fig:multiagentEx} shows the agents' trajectories
under~$\uadp$ with the choices $\delta = 0.5$, $c_\alpha = 1$,
$c_M = 100$, and $q_\alpha = q_M = 0.1$.  The plots show that agents
travel safely from the initial positions to the ending positions. The
large value of $c_M$ enhances the ability of the agents to transition
from the componets $\Cc_{i,2}$, where they start, to the components
$\Cc_{i,1}$, where they end. Small values of $c_M$ might result in
agents not leaving their respective $\Cc_{i,2}$'s, i.e., safety would
still be ensured but not the control objective of getting the agents
to their final destinations.
% This observation is important as it highlights that our design here
% ensures safety, potentially at the cost of sacrificing other control
% objectives.  Since we knew that our other control objectives in this
% example require transisiton from one component of the safe set to the
% other, we applied a large $c_M$ to promote that, and thus ended up
% with safety not interfering with other objectives.
%
% \marginJC{It's a bit sad to end the paper like this. It's better if
%   you take the time to wrap it up properly: what does the figure show:
%   successful, unsuccesful execution. Anything of note? Any interesting
%   insights the reader might not notice unless we draw attention to
%   them? What happens to executions if one plays with the parameter
%   choices $c_\alpha$, $c_M$, $q_\alpha$, $q_M$? In other words, what's
%   the impact of larger/smaller $\alpha$, $M$ (see earlier margins)?}
%\marginMA{Athough not very happy.. it is now less sad.}
%

\section{Conclusions}

We designed controllers that render discontinuous
dynamics forward invariant with respect to non-smooth sets.  The safe
set is represented by arbitrarily nested unions and intersections of
0-superlevel sets, termed components, of differentiable functions,
resulting in a nonsmooth CBF.  We showed that the satisfaction of
the safety condition for the active components of the non-smooth CBF
does not render the discontinuous dynamics safe. We remediated
this problem by enforcing a novel version of the safety constraints
for all the components of the non-smooth CBF that incorporates
transition functions.  These functions allow system trajectories to
leave a component of the safe set to transition to a different one.
The resulting all-components QP controller is feasible, safe, and
continuous wherever the system dynamics is continuous. To enhance its
implementability, we proposed an extension termed all-components
adaptive QP controller which determines important design parameters in
an adaptive fashion.
% This representation of the safe set as nested unions and intersections
% goes beyond the state of the art in the literature which mainly works
% with safe sets represented by intersections.
Our results suggest the possibility of combining multiple safe set design
methods to achieve control objectives that might not be achievable with
one method alone.  Future work will identify conditions to ensure that the
assumption about the existence of safe control actions holds for
general system classes, extend our controller design method to
deal with time-varying safety requirements, and exploit the
flexibility of transitioning between components of the safe set to
allow richer specifications combining safety, motion planning, and
control.
% Also, our work here builds a framework in which safety critical
% control can be systematically designed to unify the many scattered
% works in the literature dedicated to approximate invariant sets.

{\small
\bibliography{../bib/alias,../bib/Main-add,../bib/JC}
\bibliographystyle{plainnat}
}
\appendix
\section{Upper Semi-continuity of Activity Set-Valued Maps}

Here we prove the upper semi-continuity of the activity set-valued maps
$\Ic$ and $\Jc$ defined in Section~\ref{sec:problem}.

\begin{lemma}\label{lem:ij}{\rm \textbf{(Upper Semi-continuity of
      Activity Set-Valued Maps).}}
  Consider the structure of the dynamics~\eqref{eq:affSys} and the
  function $h$ in~\eqref{eq:h}.
  \begin{enumerate}[(a)]
  \item For all $x \in \Xc$, there is a
    neighborhood $\Mc_x$ such that $\Ic(y) \subseteq \Ic(x)$ and
    $\Jc(y) \subseteq \Jc(x)$, for all $y \in \Mc_x$.
  \item Given the normal form of $h$
    in~\eqref{eq:hNormalForm}, let
    $L(x) \coloneqq \{\ell \in \Lc \;|\; h(x) = h^\ell(x)
    % \coloneqq \min_{i \in \Ic^\ell}h_i(x)
    \}$. Then, the neighborhood $\Mc_x$ can be taken such that
    $L(y) \subset L(x)$ for all $y \in \Mc_x$.
  \end{enumerate}
\end{lemma}
\begin{pf}
  Let $x \in \Xc$.
\begin{enumerate}[(a)]
\item If $i' \notin \Ic(x)$, then
  $x \notin \closure{\Xc^{h_{i'}}}$ by definition. Thus, there exists
  a neighborhood $B_\epsilon(x)$ such that
  % $y \notin \Xc^{h_{i'}}$ for all $y \in B_\epsilon(x)$. By the
  % openness of $B_\epsilon(x)$, 
  $y \notin \closure{\Xc^{h_{i'}}}$ for all $y \in B_\epsilon(x)$. An
  analogous argument shows that if $j' \notin \Jc(x)$, then there
  exists a neighborhood $B_{\epsilon'}(x)$ such that
  $y \notin \closure{\Xc_j}$ for all $y \in B_{\epsilon'}(x)$.  Taking
  $\Mc_x$ as the finite intersection of these neighborhoods for all
  $i' \notin \Ic(x)$ and $j' \notin \Jc(x)$ proves the statement.
\item If $\ell \notin L(x)$, then $h^\ell(x) \neq h(x)$. By continuity
  of $h$ and $h^\ell$, there exists a neighborhood where
  $h(y) \neq h^\ell(y)$ for every point $y$ in that
  neighborhood. Taking the intersection of this finite number of
  neighborhoods for all $\ell \notin L(x)$ proves the statement.
  \oprocend
\end{enumerate}
\end{pf}

\begin{lemma}{\rm \textbf{(Upper Approximation of Activity
      Set).}}\label{lem:approxI}
  Under Assumpion~\ref{as:compactC}, defining the set-valued map
  $\tilde \Ic(x)$ as in Theorem~\ref{thm:u2feasible} yields
  $\Ic(x) \subseteq \tilde \Ic(x)$ for all $x \in \Xc$.
\end{lemma}
\begin{pf}
  We reason by contradiction.  Assume there exists $x \in \Xc$ and
  $i' \in \until{r}$ such that $i' \in \Ic(x)$ and
  $i' \notin \tilde \Ic(x)$.  Thus, $h_{i'} (x) = h(x)$ and for all
  $\ell' \in \Lc_{i'}$, $h^{\ell'}(x) \neq h(x)$. By continuity, there
  exists a neighborhood $B_\epsilon(x)$ where $h^{\ell'}(y) \neq h(y)$
  for all $y \in B_\epsilon(x)$ and all $\ell' \in \Lc_{i'}$. Since
  $x \in \closure{\Xc^{h_{i'}}}$, $B_\epsilon(x) \cap \Xc^{h_{i'}}$ is
  not empty and open (since it is an intersection of two open sets),
  and thus has positive measure. Since for each
  $y \in B_\epsilon(x) \cap \Xc^{h_{i'}}$, $h(y) = h^\ell(y)$ for some
  $\ell \notin \Lc_{i'}$, there is $i \neq i'$ such that
  $h(y) = h_i(y)$.  At the same time, by definition of $\Xc^{h_{i'}}$,
  $h(y) = h_{i'}(y)$ in $B_\epsilon(x) \cap \Xc^{h_{i'}}$.
  Consequently, and given that the set $\until{r}$ is finite, there
  exists $i \neq i'$ such that $h_i(y) = h_{i'}(y)$ on a set of
  positive measure.
    % So the set of $y$ such that for
    % any $i \neq i'$, $h_i(y) = h_{i'}(y)$ has positive measure. 
    This
    contradicts Assumption~\ref{as:compactC}. 
  \end{pf}
\end{document}